\documentclass[a4paper,12pt]{article}
\usepackage[margin=1.2in]{geometry}
\usepackage{graphicx}
\usepackage{subfigure}
\usepackage[square,numbers]{natbib}
\usepackage{lipsum}
\usepackage{caption}
\usepackage{pdfpages}
\usepackage{amsmath}
\usepackage{multicol}
\usepackage{multirow}
\usepackage{float}
\usepackage{url}
\usepackage{appendix}
\usepackage{bm}
\usepackage{amssymb}

\usepackage{mwe}

\usepackage{fullpage}

\usepackage{soul, url}

\newcommand{\real}[1]{\mathbb{R}({#1})}
\newcommand{\imag}[1]{\mathbb{I}({#1})}

\newcommand{\alphaB}{\alpha_{\rm B}}

  \newcommand{\ws}{w_{\rm s}}
  \newcommand{\vs}{v_{\rm s}}

  \newcommand{\V}{\Upsilon}
  \newcommand{\Vr}{\V_{\rm ref}}

\newcommand{\uuf}{\mathbf{u}_{\rm f}}
\newcommand{\uue}{\mathbf{u}_{\rm e}}
\newcommand{\uus}{\mathbf{u}_{\rm s}}
\newcommand{\w}{\mathbf{w}}

\newcommand{\Tf}{\mathbf{T}_{\rm f}}

\newcommand{\mufr}{\mu_{\mathrm{fr}}}
\newcommand{\mue}{\mu_{\mathrm{e}}}
\newcommand{\lambdae}{\lambda_{\mathrm{e}}}
\newcommand{\rhoe}{\rho_{\mathrm{e}}}
\newcommand{\cpe}{c_P}
\newcommand{\cse}{c_S}

\newcommand{\kappafr}{\kappa_{\mathrm{fr}}}
\newcommand{\kappaf}{\kappa_{\mathrm{f}}}
\newcommand{\kappas}{\kappa_{\mathrm{s}}}
\newcommand{\rhoa}{\rho_{\mathrm{a}}}
\newcommand{\rhos}{\rho_{\mathrm{s}}}
\newcommand{\rhof}{\rho_{\mathrm{f}}}

\begin{document}

\title{Monitoring of water volume in a porous reservoir using
seismic data: Validation of a numerical model with a field experiment}

\author{M. Khalili${}^{a}$, B. Brodic${}^{f}$, P. G\"oransson${}^{c}$, S. Heinonen${}^{g}$, J.S. Hesthaven${}^{b}$, \\ A. Pasanen${}^{d}$,
M. Vauhkonen${}^{a}$, R. Yadav${}^{e}$,  and T. L\"ahivaara${}^{a}$\\
\normalsize ${}^{a}$Department of Technical Physics, University of Eastern Finland, Kuopio, Finland\\
\normalsize ${}^{b}$Computational Mathematics and Simulation Science,\\
\normalsize Ecole Polytechnique F\'ed\'erale de Lausanne, Lausanne, Switzerland\\
\normalsize ${}^{c}$Department of Engineering Mechanics,\\
\normalsize KTH Royal Institute of Technology, Stockholm, Sweden\\
\normalsize ${}^{d}$Geological Survey of Finland, Finland\\
\normalsize ${}^{e}$Nokia Solutions and Network, Finland\\
\normalsize ${}^{f}$LIAG Institute for Applied Geophysics, Hannover, Germany\\
\normalsize ${}^{g}$Institute of Seismology, University of Helsinki, Helsinki, Finland
}

\maketitle

\section*{Abstract}

As global groundwater levels continue to decline rapidly, there is a growing need for advanced techniques to monitor and manage aquifers effectively. This study focuses on validating a numerical model using seismic data from a small-scale experimental setup designed to estimate water volume in a porous reservoir. Expanding on previous work with synthetic data, we analyze seismic data acquired from a controlled experimental site in Laukaa, Finland. By employing neural networks, we directly estimate water volume from seismic responses, bypassing the traditional need for separate determinations, for example, of reservoir water-table level and porosity. The study models wave propagation through a coupled poroviscoelastic-viscoelastic medium using a three-dimensional discontinuous Galerkin method. The proposed methodology is validated against experimental data, aiming to improve precision in mapping current water volumes and contributing to the development of sustainable groundwater management practices.

\newpage

\section{Introduction}

Groundwater aquifers are facing unprecedented threats, with levels
decreasing at alarming rates, often more than one meter per year in
some areas—leading to significant long-term reductions. As a result,
surface water flows that were previously sustained by groundwater are
becoming seasonal or disappearing completely
\cite{giordano2009global}. To ensure sustainable water extraction, it
is crucial to have a better understanding of the location and extent
of groundwater resources. Motivated by these challenges, this study
explores the potential to estimate water volume in an artificial
porous reservoir from seismic measurements using a combination of
numerical modeling and neural networks, and validates the approach
using controlled field data.  Neural networks have been increasingly
applied to enhance the analysis of seismic data, for a recent review,
see \cite{mousavi}. By using neural network techniques, we aim to
directly estimate water volume from seismic responses, avoiding the
traditional need for separate determinations, for example, of
reservoir water-table level and porosity.

Geophysical methods, including seismic techniques, are commonly used
tools in the early stages of groundwater exploration and for ensuring
sustainable extraction strategies \cite{rubin2005stochastic,
  grelle2009seismic, gallardo2019hydrogeological}. They offer a
cost-effective alternative to drilling, providing laterally continuous
data across vast areas, by utilizing variations in material properties
to detect subsurface features. Seismic methods, in particular, are
well-suited for locating and monitoring water resources due to the
higher seismic velocities exhibited by saturated materials compared to
unsaturated ones \cite{kearey2002introduction}. The high resolution
of seismic methods, both horizontally and vertically, enables detailed
subsurface feature mapping.

Building upon previous works with synthetic data in both two (2D)
\cite{Lahivaara:2019eg} and three (3D) spatial dimensions
\cite{khalili2022monitoring}, this study focuses on the estimation of
water volume from seismic data collected at a controlled experimental
site in Laukaa, Finland. The seismic data, obtained from an artificial
porous sand pool using a drop-weight seismic source, were gathered
during several acquisition campaigns following controlled changes in
the water-table level. For the neural network-based water volume
estimator, we first build a synthetic training database by simulating
seismic wave propagation for different scenarios of the studied sand
pool, including variations in physical parameters and water table
level.  The simulations employ a numerical wave propagation solver
based on the methods presented in \cite{DudleyWard:2017dp,
  ward2020discontinuous, khalili2022monitoring}. This synthetic data
is then used to train neural networks, which are subsequently applied
to the real seismic data to directly recover water volume, aiming to
improve precision in mapping current water volumes and contributing to
the development of sustainable groundwater management practices.

In our analysis of neural network-based estimates, we augment our
approach by applying the Shapley additive explanation (SHAP) framework
\cite{Lundberg:2017ua}. Employing the SHAP framework provides us with
a deeper understanding of the estimation process at the receiver
level. Specifically, we use SHAP analysis to evaluate the contribution
of each receiver to the final water-volume estimate. The purpose is
interpretability rather than receiver-array optimization.

In this study, we present a comprehensive framework for estimating
water volume in a porous reservoir using seismic data and neural
networks, validated through field experiments at the aforementioned
Laukaa test site. Section~\ref{sec:meas2} provides an overview of the
site and details the seismic measurements conducted under varying
water table conditions. In Section~\ref{sec:for}, we describe the
construction of a 3D numerical model that replicates the test site,
used for simulating synthetic seismograms based on Biot’s
poroviscoelastic theory. Section~\ref{sec:inv} introduces the neural
network architecture, including the data preprocessing steps, training
methodology, and interpretability analysis using SHAP values. Results
are shown in Section~\ref{sec:results}, with a focus on the water
volume predictions and the SHAP analysis. In Section \ref{sec:conc},
we summarize our findings.

\section{Site description and seismic measurements}\label{sec:meas2}

The field measurements were conducted in a man-made sand pool at the
Natural Resources Institute Finland (Luke) in Laukaa,
Finland. Generally, the sand pool serves as a controlled environment
to explore groundwater distribution with a knowledge of the media’s
geometry and physical parameters. Although natural small-scale
variability exists, the site can be reasonably approximated as a
homogeneous and isotropic sand body for the purposes of our
modelling. For conceptual clarity, we divide the domain into three
approximate zones: an air-saturated porous region, a water-saturated
porous region, and the surrounding elastic material. The pool is
enclosed by an impermeable clay lining \cite{pulkkinen2021start} and
its dimensions are well defined, allowing controlled adjustments of
the water-table level. This makes it possible to compute the true
water volumes for each measurement configuration. The water-saturated
region occupies the bottom of the pool (23.5 m by 8.2 m at 2 m depth),
while the overlying air-saturated zone extends to approximately 29.5 m
by 14.2 m.

The seismic measurements were conducted in June 2022. In our
experiments, a metallic rod weight was dropped on a steel plate and
served as a source of seismic waves, with an electric brake mechanism
to control the rod's release and prevent multiple hits. The
experiments involved shots at 13 specific locations and were conducted
at three different drop heights ($\rm{dh}$): 5, 10, and 15 cm. The
different drop heights were used to vary the source amplitude and
check for potential signal clipping at the nearest receivers.

For data acquisition, three-component (3C) 5 Hz geophones connected to
24-bit nodal seismic recorders were used, obtained from the Finnish
Seismic Instrument Pool (FINNSIP) \cite{hillers25}. Various sources
of ambient noise were present during the field campaign, including
river currents, nearby construction activity, and occasional lawnmower
operation. To reduce their influence, measurements were scheduled
during rain-free and relatively quiet periods. The instruments were
deployed along four receiver lines (14 receivers per line) with
approximately 1.0 m inline and 1.7 m crossline spacing, plus one
additional station placed at a random location, forming a fine-scale
3D acquisition layout.  Exact receiver locations were surveyed using
Differential Global Positioning System (DGPS) and incorporated
directly into the numerical simulations. The seismic data were
recorded at 4 kHz sampling rate.

\begin{figure}[!htb]
\includegraphics[width=0.7\textwidth]{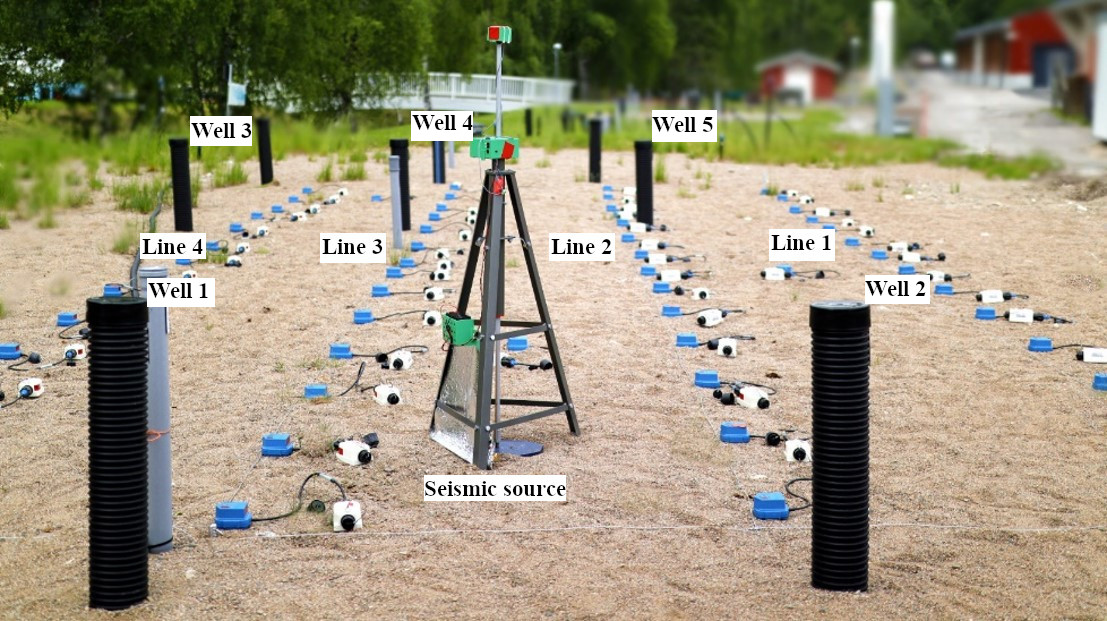}
\centering
\caption{Photo of experimental setup at Laukaa test site, showing the grid of geophones,
weight-drop source, and wells to measure the water-table level.} \label{fig:field}
\end{figure}

Figure \ref{fig:field} shows the Laukaa test site, including the
layout of the geophones, the weight-drop source, and the black pipes
housing the water-table gauge wells. A total of 57 three-component
geophones (blue) and their corresponding nodal recorders (white boxes)
were deployed.  To obtain ground-truth water-volume estimates,
water-table levels were monitored using these gauge wells at several
locations around the pool. The water level was adjusted by releasing
water to the adjacent river, after which the system was allowed to
stabilize before seismic acquisition. These auxiliary measurements
were used exclusively for validating the estimated water volumes.

\section{Modeling of the test site and field measurements}\label{sec:for}

This section presents a three-dimensional (3D) synthetic model
developed to simulate wave propagation and generate training data for
the neural network. The model's geometry for synthetic data generation
replicates the Laukaa sand pool, capturing its structure and physical
parameters. This replicated model serves as a tool for generating
extensive synthetic training data for our neural network model. To
generate synthetic seismograms, we employ Biot’s isotropic
poroviscoelastic model to handle wave propagation in the porous
medium. In a zone adjacent to the porous material, the isotropic
viscoelastic model is utilized.

\subsection{Governing equations}\label{sec:governing}

We model wave propagation in porous and elastic media using well-established partial differential equations (PDEs). Here, we summarize the main PDE systems and the associated physical parameters. More detailed derivations, numerical implementations, and attenuation models are available in \cite{khalili2022monitoring} and references therein.

\subsubsection{Biot's poroelastic wave equation}

The coupled dynamics of solid and fluid phases in porous media are described by Biot's equations \cite{carcione15}
\begin{align}
  \rhoa \frac{\partial^2 \uus}{\partial t^2} + \rhof \frac{\partial^2 \w}{\partial t^2} &= \nabla \cdot \mathbf{T}, \label{eq:biot_solid}\\
  \rhof \frac{\partial^2 \uus}{\partial t^2} + m \frac{\partial^2 \w}{\partial t^2} + \frac{\eta}{k} \frac{\partial \w}{\partial t} &= \nabla \cdot \Tf, \label{eq:biot_fluid}
\end{align}
where $\uus$ is the solid displacement, and $\w = \phi(\uuf - \uus)$
is the relative fluid displacement (scaled by porosity $\phi$). Here,
$\uuf$ denotes the fluid displacement. The average density is $\rhoa =
(1-\phi)\rhos + \phi \rhof$, where $\rhof$ is the fluid density,
$\rhos$ is the solid density, $\eta$ is the fluid viscosity, and $k$
is the permeability. The parameter $m = \rhof \tau / \phi$, with
tortuosity $\tau$.

The total stress tensor $\mathbf{T}$ and fluid stress tensor $\Tf$ are given by
\begin{align}
  \mathbf{T} &= 2\mufr \mathbf{E} 
    + \big(\lambda \operatorname{tr}(\mathbf{E}) - \alphaB M \zeta \big) \mathbf{I}, \\
  \Tf &= \big(\alphaB M \operatorname{tr}(\mathbf{E}) - M \zeta \big) \mathbf{I},
\end{align}
where $\mathbf{E}$ is the solid strain tensor and $\zeta = -\nabla
\cdot \w$ is the fluid content variation. The parameter $\mufr$ is the
shear modulus of the frame, and $\mathbf{I}$ denotes the identity
tensor.

The remaining parameters, $\lambda$, $\alphaB$, and $M$, are defined as
\begin{align}
  \lambda &= \kappafr + \alphaB^2 M - \frac{2}{3} \mufr, \\
  \alphaB &= 1 - \frac{\kappafr}{\kappas}, \\
  M &= \frac{\kappas}{\alphaB - \phi(1 - \kappas / \kappaf)},
\end{align}
where $\kappafr$, $\kappas$, and $\kappaf$ are the frame, solid, and fluid bulk moduli, respectively.

\subsubsection{Elastic wave equation}

For purely elastic media, the displacement $\uue$ satisfies
\begin{equation}
  \rhoe \frac{\partial^2 \uue}{\partial t^2} = \nabla \cdot \mathbf{S},
\end{equation}
with stress tensor
\begin{equation}
  \mathbf{S} = 2 \mue \mathbf{E} + \lambdae \operatorname{tr}(\mathbf{E}) \mathbf{I},
\end{equation}
where $\rhoe$ is the density, and $\mue$, $\lambdae$ are the Lamé parameters. These parameters determine the compressional and shear wave speeds
\begin{equation*}
  c_P = \sqrt{\frac{\lambdae + 2\mue}{\rhoe}}, \quad c_S = \sqrt{\frac{\mue}{\rhoe}}.
\end{equation*}

At interfaces between elastic and poroelastic media, we follow the
coupling strategy used in \cite{DudleyWard:2017dp,
  ward2020discontinuous}. The interface states are obtained by solving
the corresponding elastic–poroelastic Riemann problem, subject to the
physically required continuity conditions: continuity of solid
velocity (normal and tangential) and continuity of total traction,
while the normal fluid velocity vanishes on the elastic side.

\subsubsection{Modeling attenuation}

Attenuation in elastic and poroelastic media arises from different physical mechanisms. In both cases, stiffness attenuation is incorporated using a generalized Maxwell body (GMB) rheology \cite{moczo2014finite}. The GMB introduces additional free parameters, specifically relaxation frequencies and memory variables, which govern the temporal behavior of stress relaxation and allow frequency-dependent attenuation. In elastic media, these parameters are tuned to match the desired quality factors $Q_P$ and $Q_S$. In poroelastic media, the quality factors $Q_{\mufr}$, $Q_{\kappafr}$, $Q_{\kappas}$, and $Q_{\kappaf}$ are included in the model to account for frequency-dependent attenuation of the frame, solid, and fluid components. The full formulation of the GMB-based attenuation model, including how these memory variables are coupled with the governing PDEs and how the unrelaxed material parameters are derived, is provided in \cite{khalili2022monitoring} and is not repeated here. For completeness, we note that the attenuation formulation used in \cite{khalili2022monitoring} follows the poroelastic and viscoelastic GMB developments presented in \cite{zhan2019complete,7956216}.

In poroelastic media, viscodynamic attenuation also plays a role, arising from viscous drag between fluid and solid phases. In this work, we operate within Biot’s low-frequency regime, where the dominant dissipation mechanism is frequency-independent fluid flow resistance. This results in a stiff PDE system due to the large contrast between relaxation and wave propagation timescales. We express the governing equations in first-order form using a velocity-strain formulation, leading to a hyperbolic system with stiff source terms. To handle this numerically, we apply Godunov splitting \cite{leveque02} to separate the dissipative and conservative components. Following \cite{shukla2019}, we solve the stiff dissipative terms analytically at each time step, while integrating the remaining hyperbolic system using the wave propagation software described in Section~\ref{sec:dg}.

\subsection{Synthetic model of the Laukaa test site}\label{sec:prob}

\begin{figure}[!htb]
\includegraphics[width=0.79\textwidth]{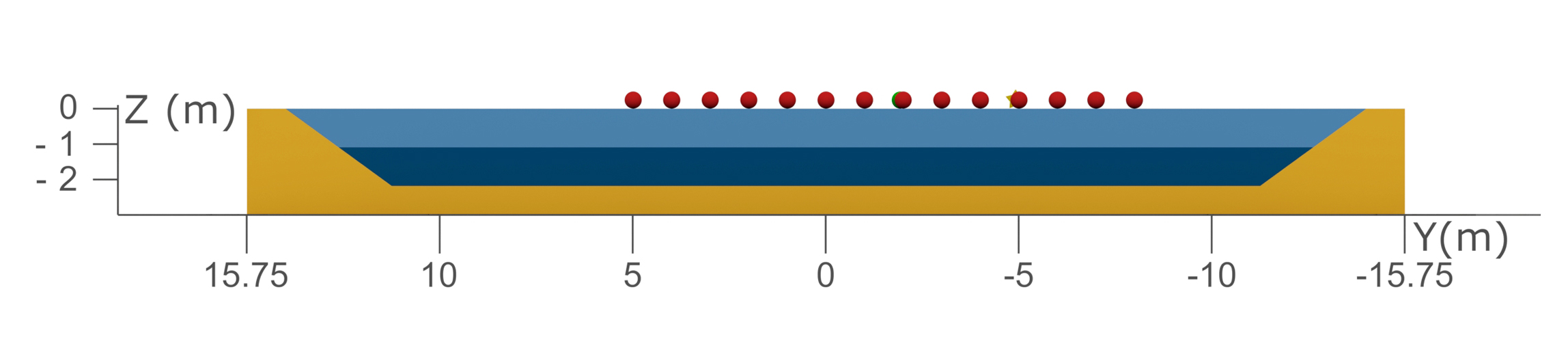}
\includegraphics[width=0.79\textwidth]{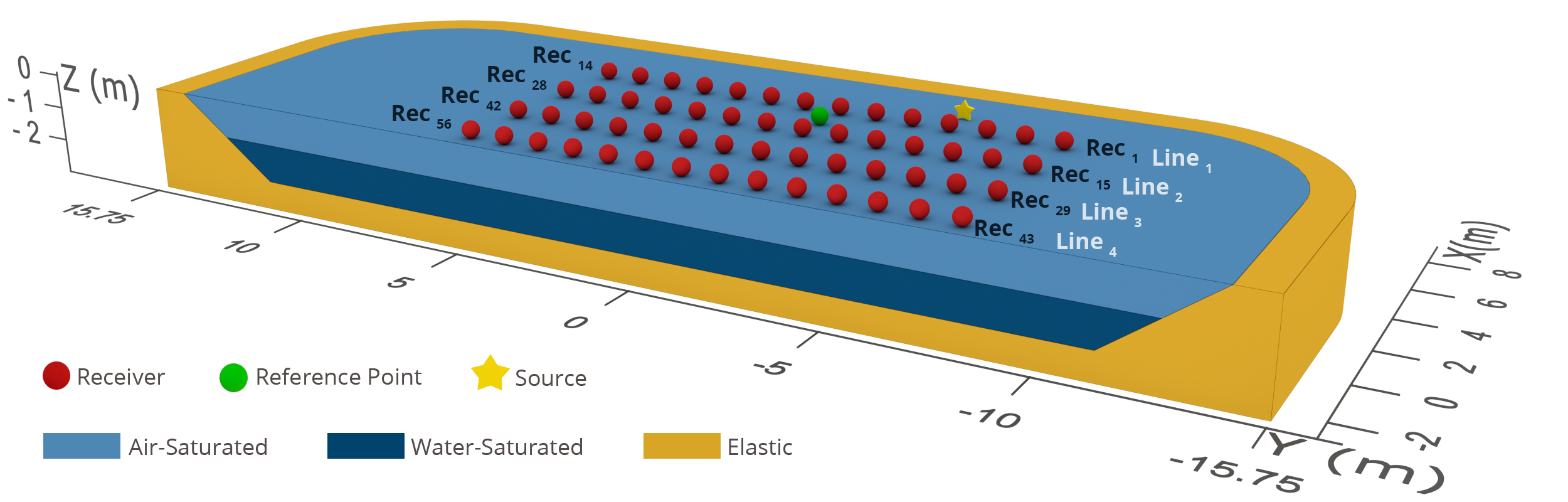}
\centering
\caption{A schematic of the problem geometry with the top showing cross-section and the bottom an oblique angle 3D view. Light-blue colour refers to air-saturated and dark-blue to water-saturated zone. Light-brown colour denotes the surrounding elastic material. The setup contains a total of 57 receivers, marked with red dots and one green dot, all of which are located on the ground surface. The green dot serves as the reference receiver and is located between the seventh receiver on lines 1 and 2. The source location is marked with a yellow star and placed close to the fourth receiver on line 1. This figure is motivated by the problem geometry presented in \protect\cite{khalili2022monitoring}.}
\label{fig:graph}
\end{figure}

The applied synthetic model represents a finite section of the porous medium used for numerical simulations. It is a rectangular box with a length of 31.5 m, a width of 16.2 m, and a height of 2.75 m, with slightly rounded corners as shown in Fig.~\ref{fig:graph}. It includes realistic air- and water-saturated subdomains, and a viscoelastic surrounding. The water-table level $z$ in the synthetic model is randomized as $z \sim \mathcal{U}(-120, -25)$ cm from the top surface.

Our model includes a total of 57 receivers that record the solid velocity components in the horizontal ($y$) and vertical ($z$) directions, denoted by $\vs$ and $\ws$, respectively. These components represent the solid velocity in the model's coordinate system (see Fig.~\ref{fig:graph}). Notably, they do not correspond to isolated wave modes (e.g., P-, SV-, or SH-waves), but rather capture the full wavefield projected onto the model axes. As a result, the signals may contain contributions from multiple wave types, due to mode conversions and the complex nature of wave propagation in porous media. One additional receiver, marked in green, is located between the lines 1 and 2 which serves as a reference point. A more detailed discussion of this reference point is provided in Section \ref{sec:matmod}.

Although the measurement data acquisition involved 13 shot locations, the water volume estimation is based on data from a single source location only, as used in the simulation study \cite{khalili2022monitoring}. The location for the source was arbitrarily selected to be the closest to receiver 4 on Line 1. The seismic source is modeled as a vertical force pointing to the negative $z$-axis. Since the exact location of the source slightly varies between different measurements (distance from the closest geophone was measured with a ruler), the location was assumed to be uncertain in the numerical model when training the neural network. In practice, to account for uncertainty in repositioning the source between measurements at different water-table levels, we assume a 10~cm variability in the source location in both the $x$ and $y$ directions. This means that the source center location in the $(x, y)$-plane is randomized as $x \sim \mathcal{U}(4.4890, 4.5890)$~m and $y \sim \mathcal{U}(-5.2142, -5.1142)$~m. Receivers and the source are placed on the ground surface. 

\subsection{Physical parameters}\label{sec:params}

The computational domain consists of three layers: an elastic
surrounding medium and two porous subdomains, which are
water-saturated and air-saturated. The elastic medium is modeled as
viscoelastic, with properties defined by solid density $\rhoe$,
pressure and shear wave speeds ($\cpe$, $\cse$), and quality factors
$Q_P$ and $Q_S$, which account for attenuation. The porous subdomains
share the same solid matrix properties, including grain density,
porosity, tortuosity, solid and frame moduli, and grain-size
distribution. All of these material parameters are randomized from
uniform distributions, with bounds provided in
Table~\ref{tab:layer_params}. The porous reservoir parameter values in
Table~\ref{tab:layer_params} are based on site-specific information
provided in the internal report (in Finnish and not publicly
available) of the Geological Survey of Finland (GTK) as well as
\cite{pulkkinen2021start, MavkoRockPhysics}. The surrounding material
(marked with light-brown in Fig.~\ref{fig:graph}) is approximated by
an idealized elastic medium whose parameters are selected based on the
original structural descriptions of the site.

The distinction between the two porous layers lies in the fluid phase:
the water-saturated domain uses fixed values for fluid density, bulk
modulus, and viscosity specific to water, while the air-saturated
domain uses different fixed values representative of air. These fluid
parameters are also listed in Table~\ref{tab:layer_params}.

Permeability $k$ is derived from the Darcy law
\begin{equation}
    \frac{k}{\eta} = \frac{K}{\rho_f g},
\end{equation}
where $g = 9.81$ m s$^{-2}$, and $K$ is the hydraulic conductivity estimated from grain-size parameters $d_{10}$ and $d_{50}$ following \cite{alyamani93}
\begin{equation}
    K\, (\mbox{m/s}) = a\left(I_0 + 0.025(d_{50} - d_{10})\right)^2,
\end{equation}
with $a = 1.505 \times 10^{-3}$, and $I_0$ denoting the extrapolated lower-limit grain size \cite{alyamani93}. The grain-size parameters are also randomized as shown in Table~\ref{tab:layer_params}. We compute the permeability value for each material sample using the viscosity and density values for water.  One must note, that the attenuation is modeled with three mechanisms in both porous reservoir and the surrounding medium, see \cite{khalili2022monitoring}. 

\begin{table}[!htb]
\centering
{
\caption{Material parameters used in the three-layer model. Layers: (1) elastic surrounding medium, (2) porous reservoir (water-saturated), (3) porous zone (air-saturated). Porous layers share the same solid matrix properties.}
\label{tab:layer_params}
\begin{tabular}{lllll}
\hline
\textbf{Layer} & \textbf{Parameter} & \textbf{Symbol (unit)} & \textbf{Min} & \textbf{Max} \\
\hline
\multicolumn{5}{l}{\textbf{Layer 1: Elastic surrounding medium (viscoelastic)}} \\
& Solid density & $\rhoe$ (kg m${}^{-3}$) & 1400 & 1800 \\ 
& Pressure wave speed & $\cpe$ (m s$^{-1}$) & 1000 & 2000 \\
& Shear wave speed & $\cse$ (m s$^{-1}$) & 400 & 800 \\
& Quality factor (P-wave) & $Q_P$ & 20 & 50 \\
& Quality factor (S-wave) & $Q_S$ & 20 & 50 \\
\hline
\multicolumn{5}{l}{\textbf{Layers 2–3: Porous matrix (shared by both saturated zones)}} \\
& Grain density & $\rhos$ (kg m${}^{-3}$) & 2400 & 2800 \\
& Solid bulk modulus & $\kappas$ (GPa) & 45 & 55 \\
& Frame bulk modulus & $\kappafr$ (GPa) & 0.008 & 0.05 \\
& Frame shear modulus & $\mufr$ (GPa) & 0.002 & 0.04 \\
& Porosity & $\phi$ (\%) & 30 & 40 \\
& Tortuosity & $\tau$ & 1.1 & 1.8 \\
& Quality factor (Shear modulus) & $Q_{\mufr}$ & 15 & 50 \\
& Quality factor (Solid bulk modulus) & $Q_{\kappas}$ & 80 & 120 \\
& Quality factor (Frame bulk modulus) & $Q_{\kappafr}$ & 15 & 50 \\
& Quality factor (Fluid bulk modulus) & $Q_{\kappaf}$ & \multicolumn{2}{c}{$\infty$} \\
& Grain size & $d_{10}$ (mm) & 0.4 & 0.8 \\
& Grain size & $d_{50}$ (mm) & 1.1 & 1.6 \\
\hline
\multicolumn{5}{l}{\textbf{Layer 2 only: Fluid parameters (water-saturated)}} \\
& Fluid density & $\rhof$ (kg m${}^{-3}$) & \multicolumn{2}{c}{1000} \\
& Fluid bulk modulus & $\kappaf$ (GPa) & \multicolumn{2}{c}{2.1025} \\
& Fluid viscosity & $\eta$ (Pa$\cdot$s) & \multicolumn{2}{c}{$1.14 \times 10^{-3}$} \\
\hline
\multicolumn{5}{l}{\textbf{Layer 3 only: Fluid parameters (air-saturated)}} \\
& Fluid density & $\rhof$ (kg m${}^{-3}$) & \multicolumn{2}{c}{1.2} \\
& Fluid bulk modulus & $\kappaf$ (MPa) & \multicolumn{2}{c}{0.13628} \\
& Fluid viscosity & $\eta$ (Pa$\cdot$s) & \multicolumn{2}{c}{$1.8 \times 10^{-5}$} \\
\hline
\end{tabular}
}
\end{table}

The volume of stored water in the reservoir can be calculated by multiplying the volume of the water-saturated domain by the porosity. In this paper, the amount of water is calculated only from the water-saturated zone that is located exactly under the array of receivers. The current problem setup, with its simplified geometric structure, allows for the potential estimation of water volume across the entire reservoir. However, this cannot be assumed to be directly applicable to more complex and realistic aquifer models. 

For the Laukaa test case studied in this paper, the parameter distributions shown in Table \ref{tab:layer_params} are chosen to be representative of the geological properties observed in the region. However, it is important to note that while these ranges are designed to capture key variations, the neural network model may still perform well even if not all parameters are completely accounted for in the training data. This is because the network is able to learn relevant features from the available data, potentially allowing it to generalize and provide accurate estimates even in regions not explicitly covered by the training set. In this sense, the model can adapt to variations in parameters and still maintain robustness in its predictions \cite{goodfellow2016deep}.

\subsection{Computation of seismic data}\label{sec:dg}

We utilize an in-house forward solver based on the discontinuous
Galerkin (DG) method \cite{Hesthaven:2007nd} and third-order
Adams–Bashforth time-stepping \cite{Durran:1991ta} to compute
synthetic seismograms. The solver is implemented in C/C++ and
parallelized across CPU nodes using the Message Passing Interface
(MPI). GPU acceleration is achieved through the OCCA library
\cite{medina2014occa}, which enables communication between the CPU
and GPU as well as efficient execution on graphics hardware. The
computational domain is discretized using tetrahedral elements, and
accuracy is controlled via element size and the order of polynomial
basis functions. All simulations were performed on the Puhti
supercomputing environment maintained by CSC -- IT Center for Science,
Finland \cite{CSC}, using NVIDIA Volta V100 GPUs. Computational grids
used in this work were built using COMSOL Multiphysics.

Two source time functions are used in our modeling: the first
derivative of a Gaussian and the Ricker wavelet. Both wavelets are
defined below using the same frequency and time delay parameters. We
employ two source signatures for the following reason: the first
derivative of a Gaussian wavelet is used to generate the training and
validation datasets (Section~\ref{sec:nns}), whereas the Ricker
wavelet is used for the synthetic test set
(Section~\ref{sec:predictions}). This intentional mismatch allows us
to assess the ability of the predictive framework to generalize to an
unseen source signature.

The first derivative of a Gaussian is defined as
\begin{equation}
\label{eq:1stderiv}
    g = \frac{(t - t_0)}{c} \exp\left(b\left((t - t_0)^2 - c^2\right)\right),
\end{equation}
where $b = -(f\pi)^2$ and $c = \sqrt{-0.5/b}$.

\begin{figure}[!htb]
\centering
\includegraphics[width=0.4\textwidth]{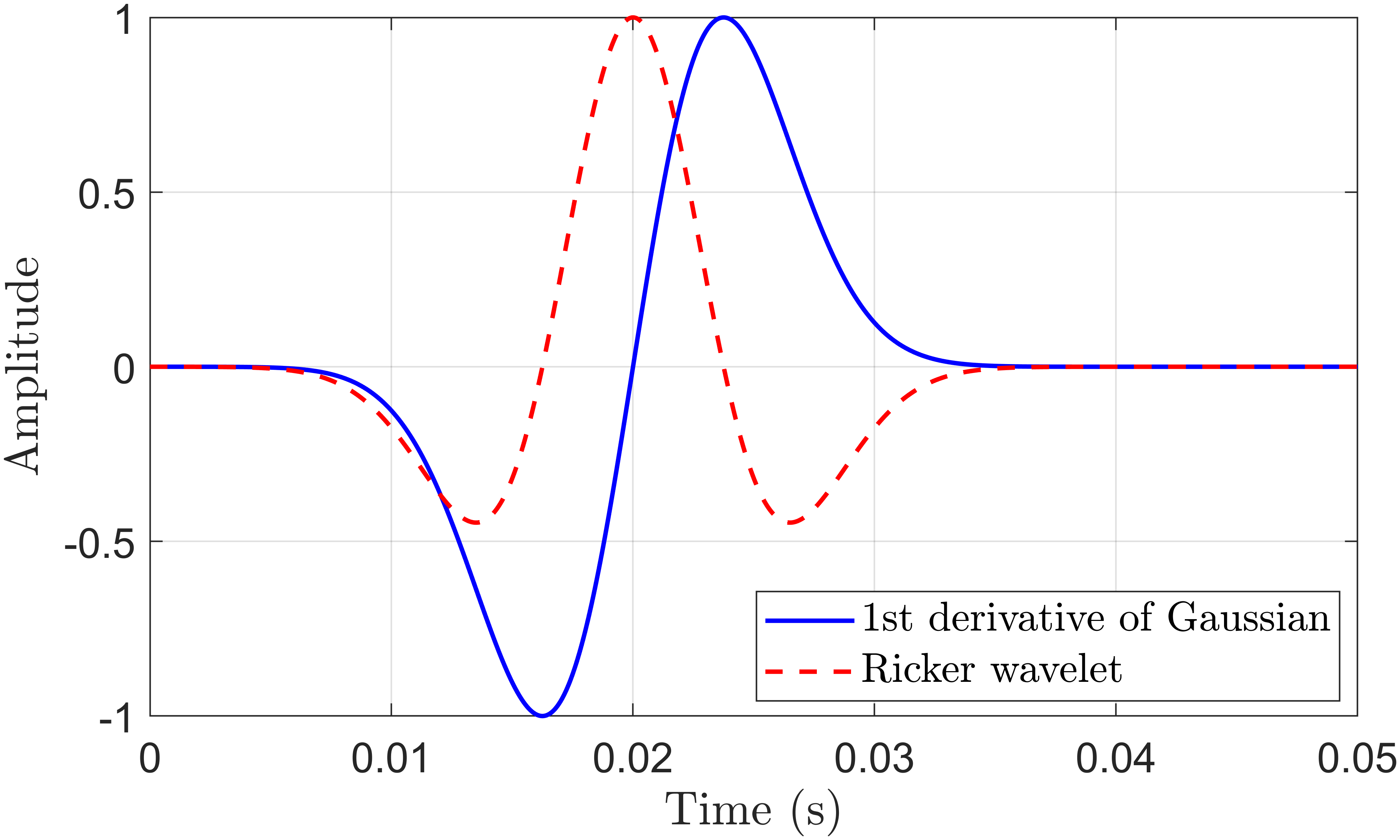}
\hfill
\includegraphics[width=0.59\textwidth]{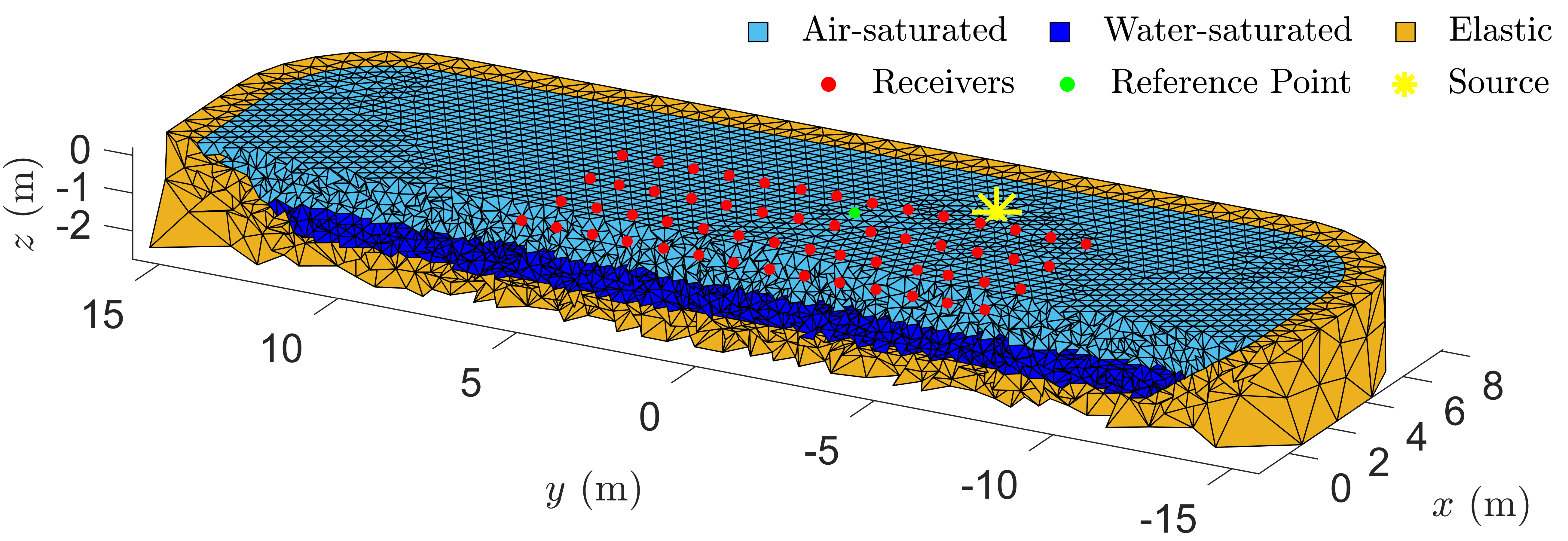}
\caption{Left: The two source time functions used in the modeling, first derivative of a Gaussian (solid blue) and Ricker wavelet (dashed red). Right: Cross section of an example mesh used in the simulations.}
\label{fig:source_mesh}
\end{figure}

The Ricker wavelet is defined as
\begin{equation}
\label{eq:2ndderiv}
    g = \left(1 + 2b(t - t_0)^2\right) \exp\left(b(t - t_0)^2\right),
\end{equation}
with the same parameters $b$ and $t_0$, using a frequency $f = 60$ Hz and delay $t_0 = 1.2/f$. Figure~\ref{fig:source_mesh}  (left) shows the source wavelet amplitudes as a function of time.

To ensure accuracy across the simulations, we use polynomial basis
functions of order four or five, depending on the dataset, and the
mesh resolution is defined relative to the smallest expected
wavelength in the model. Specifically, we use a minimum element
density criterion of approximately 2.5 elements per shortest
wavelength when generating test data (see Section
\ref{sec:predictions}), while the training and validation sets (see
Section \ref{sec:nns}) use slightly lower resolutions, approximately
2.0 and 1.9 elements per wavelength, respectively. This ensures that
the datasets have slightly different numerical accuracy, as
intended. The purpose of using different discretizations, both in mesh
resolution and basis order, is to avoid the so-called inverse crime:
if the same forward model and numerical resolution were used for
training, validation, and testing, the inversion could become
artificially easier and unrealistically accurate. This principle is
well established in the general theory of inverse problems
\cite{KaipioSomersalo}. The discretization differences introduce
controlled numerical variability. These differences are not intended
as noise but ensure that the test data are not generated with the
exact same forward operator as the training data. An example mesh used
in the synthetic test dataset simulations, shown in
Fig.~\ref{fig:source_mesh} (right), contains 158,180 tetrahedral
elements and 28,407 vertices, illustrating the level of discretization
applied.

\section{Neural network-based characterization of water storage}\label{sec:inv}

\subsection{Noise model, source function normalization, and data examples}\label{sec:matmod}

Following \cite{khalili2022monitoring} and denoting the measurement data vector and forward model as $\V=[\vs, \ws]^\top$ and $\mathcal{A}$ respectively, the observation model is given by
\begin{eqnarray}
\label{eq:obsm} 
    \V = \mathcal{A}(\boldsymbol{m}) + e = X + e, 
\end{eqnarray}
where $\boldsymbol{m}$ contains all the physical and geometrical parameters of the applied model, such as porosity, tortuosity, and water-table level. The term $e$ represents additive noise components. The forward operator $\mathcal{A}$ is used to map the model parameters $\boldsymbol{m}$ to the synthetic seismic data vector $X$, simulated using the coupled viscoelastic-poroviscoelastic material model by the DG method in three spatial dimensions \cite{ward2020discontinuous, khalili2022monitoring}. 

To simulate measurement noise, we generate the noisy trace using the model
\begin{equation}
\label{eq:noisemodel}  
X_\textrm{noised}=X+A|X|_{\max}\epsilon^A+B|X|\epsilon^B,
\end{equation}
where $\epsilon^A$ and $\epsilon^B$ represent independent zero-mean
Gaussian random variables, and $|X|_{\max}$ is the maximum absolute
value of $X$. The two noise components correspond to additive white
noise and amplitude-related noise, respectively, contributing to a
diverse range of noise levels. This decomposition is motivated by
practical considerations: in datasets with large dynamic range, a pure
white-noise model disproportionately affects weaker signals while
leaving stronger signals largely unperturbed. By including an
amplitude-related term, the noise level scales with the signal,
producing a more realistic distribution across all traces. While the
additive component can be estimated from the data, the
amplitude-dependent component cannot be robustly inferred. Therefore,
parameters $A$ and $B$ are varied over wide intervals ($A \in
[0.05,2]\%$, $B \in [0,2]\%$) to explore different noise conditions. 

The seismic measurements were recorded continuously, but each source
excitation produced a 3-second-long data segment, which was downloaded
from the recorders based on microsecond-accurate, GPS-timed source
initiations. For the neural network input, a 0.35-second window was
extracted from both field and simulated data. This duration ensures
that all significant wave reflections and late arrivals have
attenuated to negligible amplitudes, preventing artificial truncation
of energy and allowing each trace to end near zero amplitude.

After the waves produced by the seismic source have attenuated, we
estimated the standard deviation of the Gaussian white noise component
by analyzing a time window between 1 and 2 seconds. Comparing the
estimated noise amplitude to peak signal amplitudes yielded an
approximate noise level of $A_{\rm meas} \approx 0.4\%$.

In order to obtain a source-independent inversion, we use a
deconvolution operation to remove the effect of the seismic source
time function \cite{Lee:2003si}. We transform the transient signals
to the frequency domain and use data from a reference receiver (see
Fig.~\ref{fig:graph}) as the system response function. This allows us
to rewrite the observation model (Eq.~\ref{eq:obsm}) as
\begin{eqnarray}
    \label{eq:obf_comp}
    \frac{F_{\V}(\omega_\ell)}{F_{\Vr}(\omega_\ell)} &=& 
    \frac{F_X(\omega_\ell)}{F_{X_{{\rm ref}}}(\omega_\ell)} + \hat{e}_\ell,
    \qquad \ell = 1,\ldots,N_f, \\
    \label{eq:obf}
    \Leftrightarrow\quad \hat{V} &=& \hat{X} + \hat{e},
\end{eqnarray}
where $F_{\V}$ is the Fourier transform of the measured seismic data
at a given receiver, $F_{\Vr}$ is the Fourier transform of the data at
the reference receiver, $F_{X}$ denotes the Fourier transform of the
simulated (noise-free) model response, and $F_{X_{\rm ref}}$ is the
corresponding simulated response at the reference receiver. Here,
$\hat{e}$ is the noise term in the frequency-domain formulation after
normalization, $\omega_\ell$ is the frequency at index $\ell$, and
$N_f$ is the total number of frequencies. Note that the
model~(\ref{eq:obf}) is applied separately to both velocity
components, $\vs$ and $\ws$, at each receiver.

In practice, the division in Eq.~(\ref{eq:obf}) is implemented using
the conjugate of $F_{X_{\rm ref}}(\omega_\ell)$, leading to a
denominator of the form $|F_{X_{\rm ref}}(\omega_\ell)|^2$. Since this
denominator may approach zero, we apply a Wiener-type
regularization. Following \cite{wen}, the denominator $|F_{X_{\rm
    ref}}(\omega_\ell)|^2$ is replaced by $|F_{X_{\rm
    ref}}(\omega_\ell)|^2 + \epsilon \langle |F_{X_{\rm
    ref}}(\omega_\ell)|^2 \rangle,$ where $\epsilon = 0.1$ is a
stabilization constant. This regularization is applied directly within
the frequency-domain ratio in Eq.~(\ref{eq:obf}). It does not replace
Eq.~(\ref{eq:obf}) nor introduce any additional filtering procedure,
it only serves to avoid numerical instabilities when the reference
spectrum has small amplitudes.

\begin{figure*}[!htb]
\includegraphics[width=\textwidth]{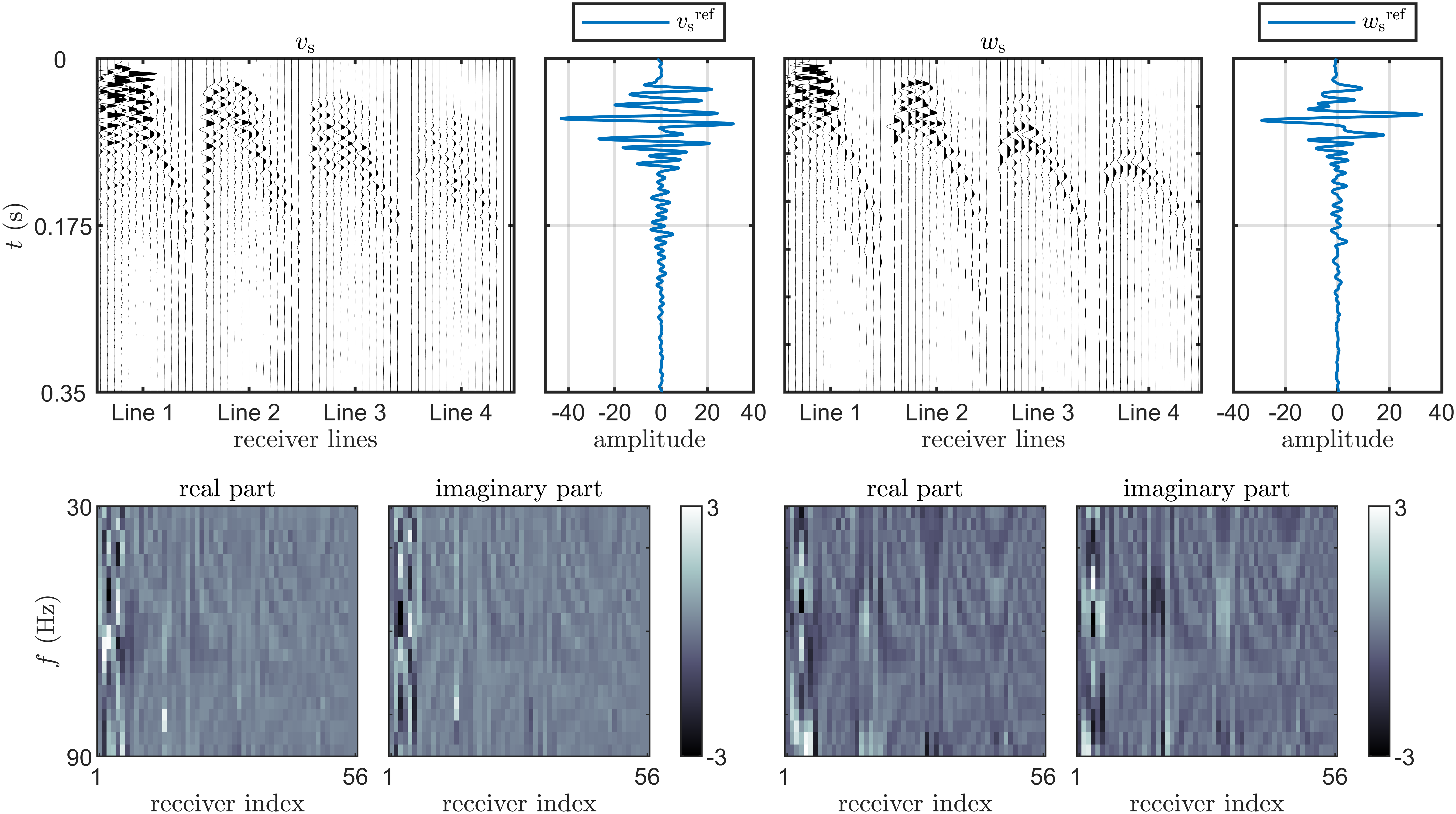}
\centering
\caption{Measured seismic data: horizontal $\vs$ component (left), vertical $\ws$ component (right). Reference traces are shown in the top right panels.}\label{fig:rdata}
\end{figure*}

To illustrate the behavior of real seismic data, Fig.~\ref{fig:rdata}
shows a sample shot gather from the field dataset, displaying the
horizontal $\vs$ and vertical $\ws$ velocity components along with
reference traces. The bottom row presents the real and imaginary parts
of the Fourier-transformed and source-normalized records, which serve
as inputs to the neural network. For this example, the water table was
at $-88.7$ cm.

\begin{figure*}[!htb]
\includegraphics[width=\textwidth]{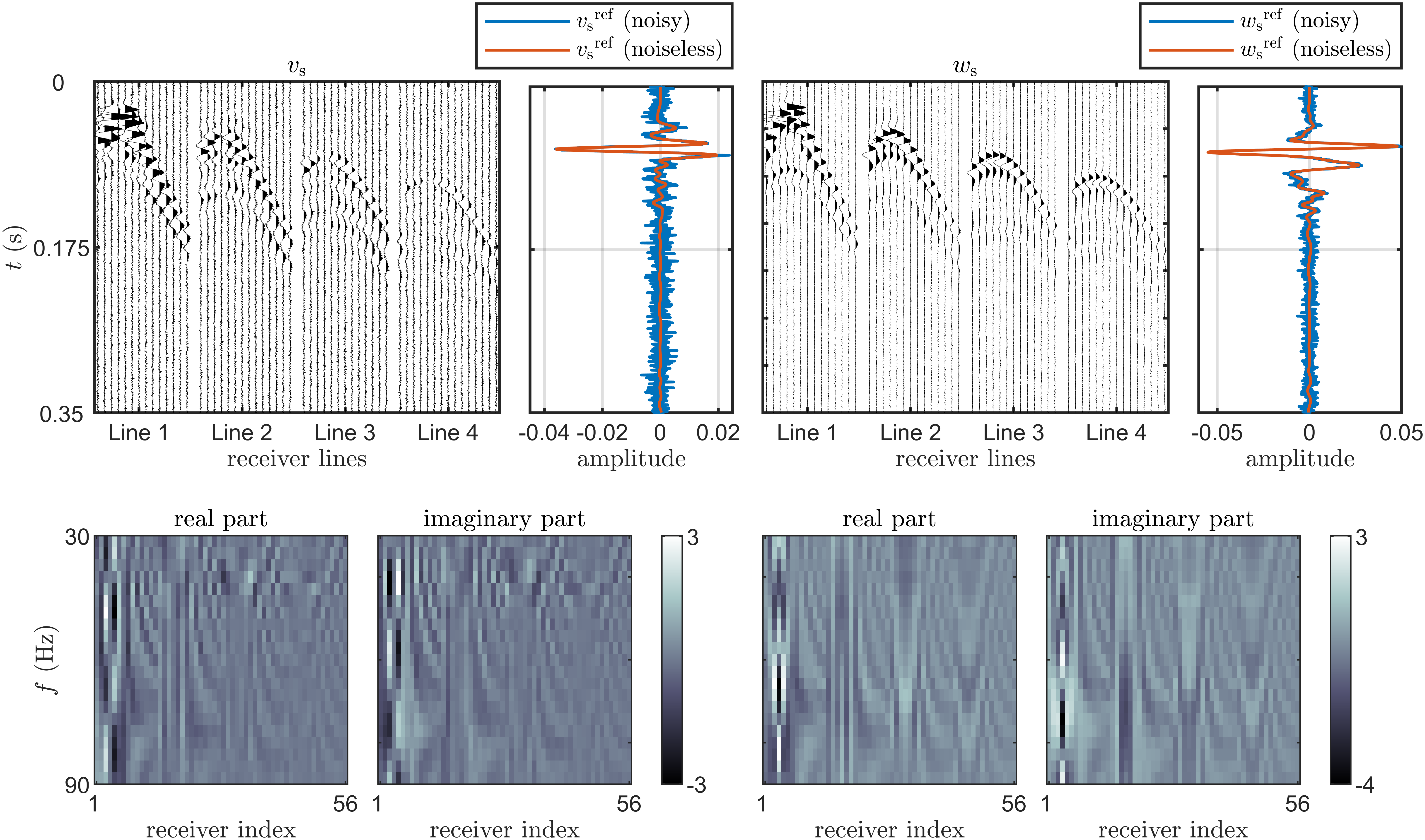}
\centering
\caption{Synthetic seismic data: horizontal $\vs$ component (left), vertical $\ws$ component (right). Reference traces for noisy and noiseless data are shown in the top panels.}\label{fig:sdata}
\end{figure*}

Similarly, Fig.~\ref{fig:sdata} presents an example of synthetic data
with a water table at $-72.10$ cm and noise parameters $A=1\%$,
$B=1\%$. Differences from real data arise from randomly selected
material parameters and differing water-table levels. These examples
illustrate the correspondence between measured and simulated records,
and the type of inputs provided to the neural network.

\subsection{Neural networks}\label{sec:nns}

We generated three distinct datasets for the neural network model: a
training set consisting of 15,000 samples and a validation set of
3,000 samples. The training and validation datasets were created using
fifth-order polynomials and the source wavelet (\ref{eq:1stderiv})
with the DG solver described in Section \ref{sec:dg}, which also
details the mesh construction and numerical discretization. The values
for the model input $\boldsymbol{m}$ were independently drawn from the
probability distributions described in Sections \ref{sec:prob} and
\ref{sec:params} and used as inputs to the forward model $\mathcal{A}$
(see Eq. (\ref{eq:obf})). In addition to training and validation
datasets, a separate test dataset comprising both synthetic and field
measurement data was employed to evaluate the final performance of the
neural network. This test dataset is described in detail in Section
\ref{sec:predictions}.

A fully connected neural network was employed in this study to
determine the water volume using seismic data. By training on data
with randomized model parameters $\boldsymbol{m}$ (see
Eq. (\ref{eq:obsm})), the network learns to focus on the relevant
seismic signatures of water volume while marginalizing the influence
of less significant factors. We down-sample the synthetic data to a
sampling frequency of 4 kHz, then generate five copies of the clean
training and validation datasets, each corrupted with Gaussian noise
according to (\ref{eq:noisemodel}). This augmentation increases the
size of both the training and validation datasets by a factor of five,
which supports better generalization and robustness to measurement
noise. The noise corrupted time domain data are transformed to the
frequency domain according to model (\ref{eq:obf}) and denoted as
$\hat{X}_\textrm{noised}$ in the following. For each sample, we
construct four input matrices of size $N_r \times N_f$, containing the
real and imaginary parts of the two velocity components. Here, $N_r =
56$ is the number of receivers and $N_f = 21$ is the number of
frequencies in the usable-energy band of the data (31.4--88.6~Hz with
$\Delta f \approx 2.86$~Hz), see Fig.~\ref{fig:sdata}. These four
matrices are stacked and reshaped into a single feature vector. This
yields an input size of
{\small
\[
\text{frequency components} \times \text{number of receivers} \times \text{velocity components} \times 2
= 21 \times 56 \times 2 \times 2 = 4704
\]
}
where the factor of 2 at the end accounts for the separate real and imaginary parts. Because the first layer is fully connected, each neuron receives all 4704 features, enabling the model to jointly learn from all four input channels. The network produces a single scalar output representing the estimated water volume.

The training process is implemented using the TensorFlow \cite{tensorflow2015-whitepaper} and Keras \cite{chollet2015keras} libraries. We employ the Adam optimizer \cite{kingma2014adam} and utilize the RandomSearch algorithm from the Keras Tuner library \cite{omalley2019kerastuner} to optimize the model’s performance by exploring different hyperparameter configurations. The Keras Tuner is allowed to randomly select the activation function, learning rate, number of hidden layers, number of neurons per layer, and $L_2$ regularization penalty factor. The activation functions considered are “relu,” “sigmoid,” “tanh,” “selu,” “swish,” and “LeakyReLU,” while the number of hidden layers ranges from one to six. The number of neurons per layer varies between 100 and 5,000. The learning rate is selected from ${1 \times 10^{-3}, 1 \times 10^{-4}, 1 \times 10^{-5}}$, and the $L_2$ regularization penalty is chosen from ${1 \times 10^{-5}, 1 \times 10^{-6}, 1 \times 10^{-7}, 1 \times 10^{-8}}$. For all networks in this study, the batch size is set to 256.

The training process is guided by minimizing the mean squared error (MSE) loss function, defined as
\begin{equation} 
\mathcal{L}(\theta) = \frac{1}{N_{{\rm train}}} \sum_{i=1}^{N_{{\rm train}}} \left(V_{{\rm true}}^{(i)} - \operatorname{NN}\left(\real{\hat{X}_\textrm{noised}^{(i)}}, \imag{ \hat{X}_\textrm{noised}^{(i)}}; \theta\right)\right)^2 + \alpha\mathcal{R}(\theta), 
\end{equation} 
where $\operatorname{NN}$ denotes the neural network with weights and biases stored in $\theta$. Additionally, $N_{{\rm train}}$ represents the number of samples in the training dataset and $V_{{\rm true}}^{(i)}$ is the true water volume for the $i$-th sample. The term $\mathcal{R}(\theta)$ denotes the sum of the squared weights of the model and $\alpha$ is the $L_2$ regularization penalty coefficient. This loss function ensures that the predicted values closely approximate the true values. The model iteratively updates its weights and biases using the Adam optimizer to minimize this loss function. The convergence of the loss function over epochs is monitored to avoid overfitting. For a broader discussion on neural network training methodologies and optimization strategies, see \cite{goodfellow2016deep, zhang2023dive}.

The best-performing network is selected based on MSE for validation. After testing various hyperparameter combinations, the final network consists of five hidden layers with 2,570 (layer 1), 3,920 (layer 2), 3,360 (layer 3), 2,730 (layer 4), and 3,400 (layer 5) neurons. It employs the LeakyReLU activation function, a learning rate of $1 \times 10^{-5}$, and an $L_2$ regularization penalty factor of $1 \times 10^{-5}$. The output layer uses a linear activation function, and early stopping is enabled during both hyperparameter tuning and training to enhance generalization and prevent overfitting.

\begin{figure*}[!htb]
\includegraphics[width=\textwidth]{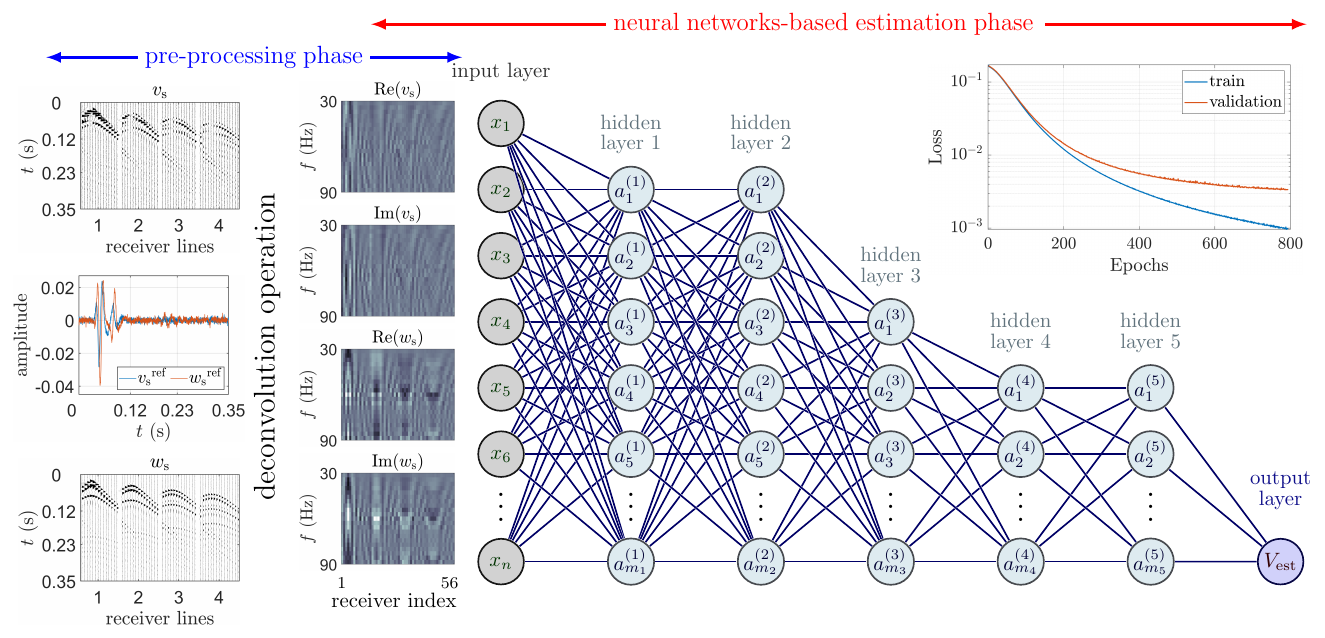}
\centering
\caption{Graphical illustration of the process for estimating water volume from seismic data. 
The left side depicts the pre-processing phase, including the deconvolution operation used to normalize the source wavelet effects. 
The right side presents the neural network-based estimation phase, showing the detailed architecture of the best-performing model. 
The network consists of five hidden layers with numbers of neurons 
$(m_1, m_2, m_3, m_4, m_5) = (2570, 3920, 3360, 2730, 3400)$. Each hidden layer uses the LeakyReLU activation function, while the output layer employs a linear activation. 
The input size to the network is $n = 4704$, corresponding to $21$ frequency components, $2$ velocity components, $56$ receivers, and $2$ for the real and imaginary parts. 
The network produces a single scalar output estimating the water volume.
The network was trained with a learning rate of $1 \times 10^{-5}$ and an $L_2$ regularization penalty factor of $1 \times 10^{-5}$. 
Early stopping was applied during both hyperparameter tuning and training to prevent overfitting and improve generalization. 
The top-right inset displays the loss curves over epochs for training and validation datasets, illustrating the model’s convergence behavior.}
\label{fig:nn}
\end{figure*}

The training process uses large, randomized datasets that broadly sample the parameter space, with noise added to help improve robustness. Hyperparameters are tuned via a random search, and early stopping is applied to reduce overfitting and support generalization to new data. Figure \ref{fig:nn} illustrates the approach for estimating water volume from seismic data. The figure shows the sequential data processing steps, the neural network architecture with five hidden layers, and the training and validation loss curves that highlight the model’s learning and generalization over time.

\section{Results}\label{sec:results}

The results are promising in terms of estimation accuracy. Notably, the estimation accuracy for the full receiver setup aligns closely with the true values and that of the supplementary synthetic database. However, it’s worth mentioning that one of the field data samples produced a significantly biased estimate when compared to the true value. Closer analysis of the data traces showed that the biased sample exhibited distinct differences, particularly in terms of the RMSE.

\subsection{Predictions of water volume}\label{sec:predictions}

We test the applicability of the trained neural network model via two different test datasets. These datasets are defined as:
\begin{itemize}
\item[] \textit{Field Measurements:} Field data were acquired at seven
  distinct water-table levels, ranging from a depth of –31.3~cm to
  –88.7~cm. At each level, three repeated measurements were recorded
  for each of three different drop heights, resulting in a dataset
  comprising $3 \times 3 \times 7 = 63$ observations. Using our
  knowledge of the actual water-table level and the geometry of the
  sand pool, we can calculate the volume of the water-saturated
  zone. Additionally, based on a previous study
  \cite{pulkkinen2021start}, we assume the nominal porosity of the
  material to be 35 per cent;
\item[] \textit{Synthetic Dataset:} This test dataset contains a total
  of 3,000 samples, distinct from the training (15,000 samples) and
  validation (3,000 samples) datasets. The same prior distributions
  were used for material parameter sampling as in the training and
  validation datasets. Noise parameters $A$ and $B$ in
  (\ref{eq:noisemodel}) were fixed to 1\% each, representing moderate
  noise levels. The DG simulations used fourth-order basis functions
  and the Ricker wavelet (see Eq.~(\ref{eq:2ndderiv})) as the source
  function. As noted earlier, a modified mesh density and polynomial
  basis order were used to introduce controlled numerical variation
  from the training and validation datasets. Additionally, the Ricker
  wavelet is employed here, differing from the first derivative of a
  Gaussian wavelet used in the training phase, to test the neural
  network’s ability to generalize to unseen source signatures.
\end{itemize}

\begin{figure}[!htb]
\includegraphics[height=0.4\textwidth]{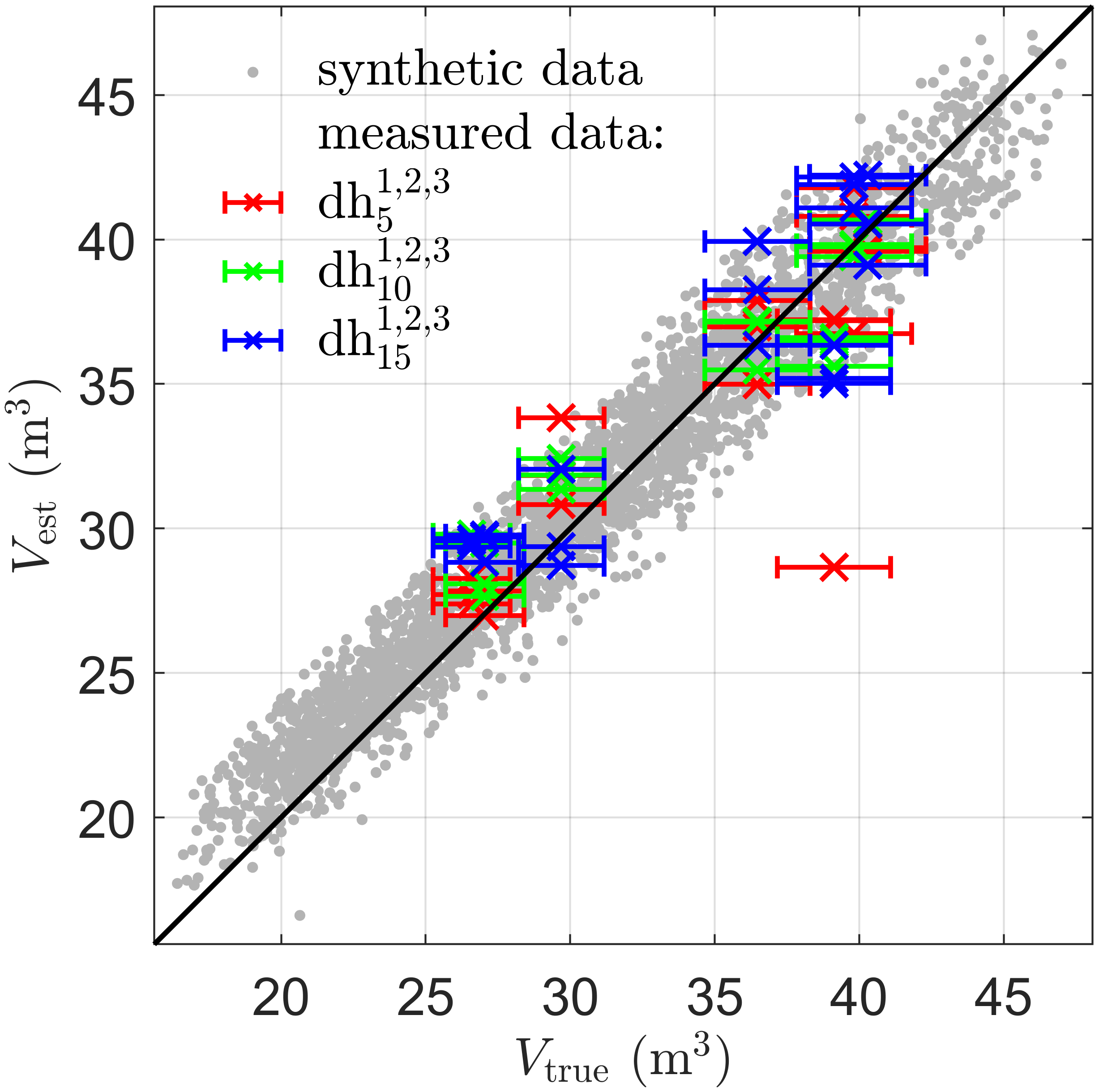}
\includegraphics[height=0.4\textwidth]{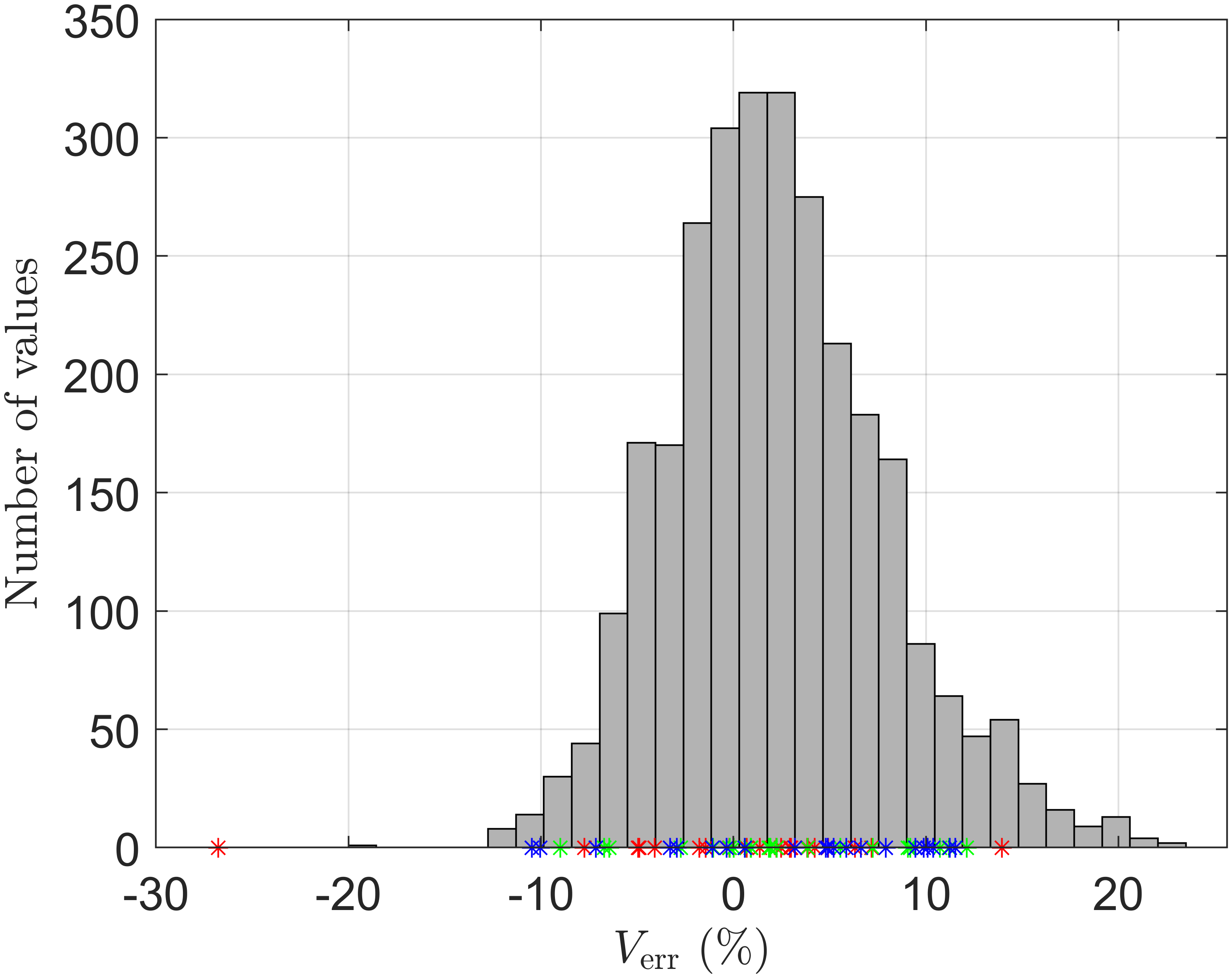}
\centering
\caption{Left: Estimated volumes of water as a function of the true value. Red denotes repeated measurements from drop height 5 cm ($\rm{dh}_{5}^{1,2,3}$), green from drop height 10 cm ($\rm{dh}_{10}^{1,2,3}$), and blue from drop height 15 cm ($\rm{dh}_{15}^{1,2,3}$).
Right: Histogram of relative prediction errors, with the colour coding of the asterisks corresponding to the same drop heights as shown on the left.} \label{fig:est}
\end{figure}

The Keras Tuner-optimized network architecture undergoes ten training
runs, and the final water volume per sample result is determined as
the average of these ten estimates. Figure \ref{fig:est} presents a
comparison between the estimated (average) water volumes and their
corresponding true values. The results corresponding to the 5\,cm,
10\,cm, and 15\,cm drop heights are shown as three distinct groups,
each containing the three repeated measurements
$(\rm{dh}_{5}^{1,2,3})$, $(\rm{dh}_{10}^{1,2,3})$, and
$(\rm{dh}_{15}^{1,2,3})$, respectively. These drop heights are colour
coded in Figure \ref{fig:est}, with red corresponding to 5 cm, green
to 10 cm, and blue to 15 cm. To get a crude approximation for the
uncertainty, we assumed that the porosity value in the field
measurements database is uncertain in a sense that we assumed
$\phi_{\rm{true}}\in[0.95, 1.05]\times35$ per cent, that can be used
to compute the error bars shown for each estimate with real data. The
figure shows also the estimates for the synthetic dataset. These
results demonstrate the potential of using proposed neural
network-based approach to recover the water volume.

These correlations are key to verifying the accuracy of our neural
network model and demonstrate its potential for practical
application. Quantitative comparison between seismic data-derived
estimates and ground truth water volumes confirms the model's
reliability under controlled conditions. However, the occurrence of a
significantly biased estimate for one field sample highlights the
challenge posed by out-of-training-distribution data, likely caused by
extreme or unexpected measurement noise, potentially hidden in the
source-to-receiver offsets. This suggests the importance of
sensitivity analysis and uncertainty quantification in further
refining the method.

The results at water-table level -36.2 cm reveal a clear outlier in Fig. \ref{fig:est}. To analyze the differences between repeated measurements, we utilize the RMSE. Let $d_i^k$ represent the normalized seismic data (see model (\ref{eq:obf})) for the $k$'th repeated measurement at drop heights $i \in \{5, 10, 15\}$ in the frequency domain, with the real and imaginary components stacked. The root mean square error (RMSE) error can now be expressed as follows:
\begin{equation}
\label{eq:rmse_data}
    {\rm RMSE}^{\ell-j}_i = \frac{\| d^{\ell}_i - d^{j}_i\|_2}{\sqrt{N}},
    \end{equation}
where $\ell$ and $j$ are the indices for repeated measurements, and $N$ is the total number of values in the data vector. Table \ref{tab:rms} lists the RMSE values between all possible combinations of input data. Specifically, the first drop height ($i = 5$) exhibits significantly larger variations compared to the other two measurements. Additionally, the combination of drop iterations $\ell=2$ and $j=3$ yields comparable RMSE values for the measurements at $i=10$ and $i=15$.

\begin{table}[!htb]
    \centering
    \caption{Root mean square errors (\ref{eq:rmse_data}) for different data combinations for measurements taken at a water-table level of -36.2 cm.}\label{tab:rms}
    \begin{tabular}{cccc}
\hline
     $i$ & $\ell=1, j=2$ & $\ell=1, j=3$ & $\ell=2, j=3$ \\
\hline
     5 &   0.3780  &  0.3781 &   0.1399 \\ 
     10 &  0.1695  &  0.2112 &   0.1162 \\
     15 &  0.1401  &  0.1591 &   0.1070 \\
\hline
    \end{tabular}
\end{table}

\subsection{SHAP analysis}

We applied Shapley Additive Explanations (SHAP) analysis to the full receiver array neural network model to determine the significance of each receiver in estimating water volume. Determining Shapley values is an attribution problem, which means it involves determining the contribution of the prediction scores of a model for a specific sample input to its base features—in our case, the receivers. In simple terms, attribution to a base feature represents the importance of that feature to the prediction. For example, when attribution is applied to a model that estimates water volume, it helps us understand how influential each receiver is in determining the water volume.

\begin{figure}[!htb]
\includegraphics[width=\textwidth]{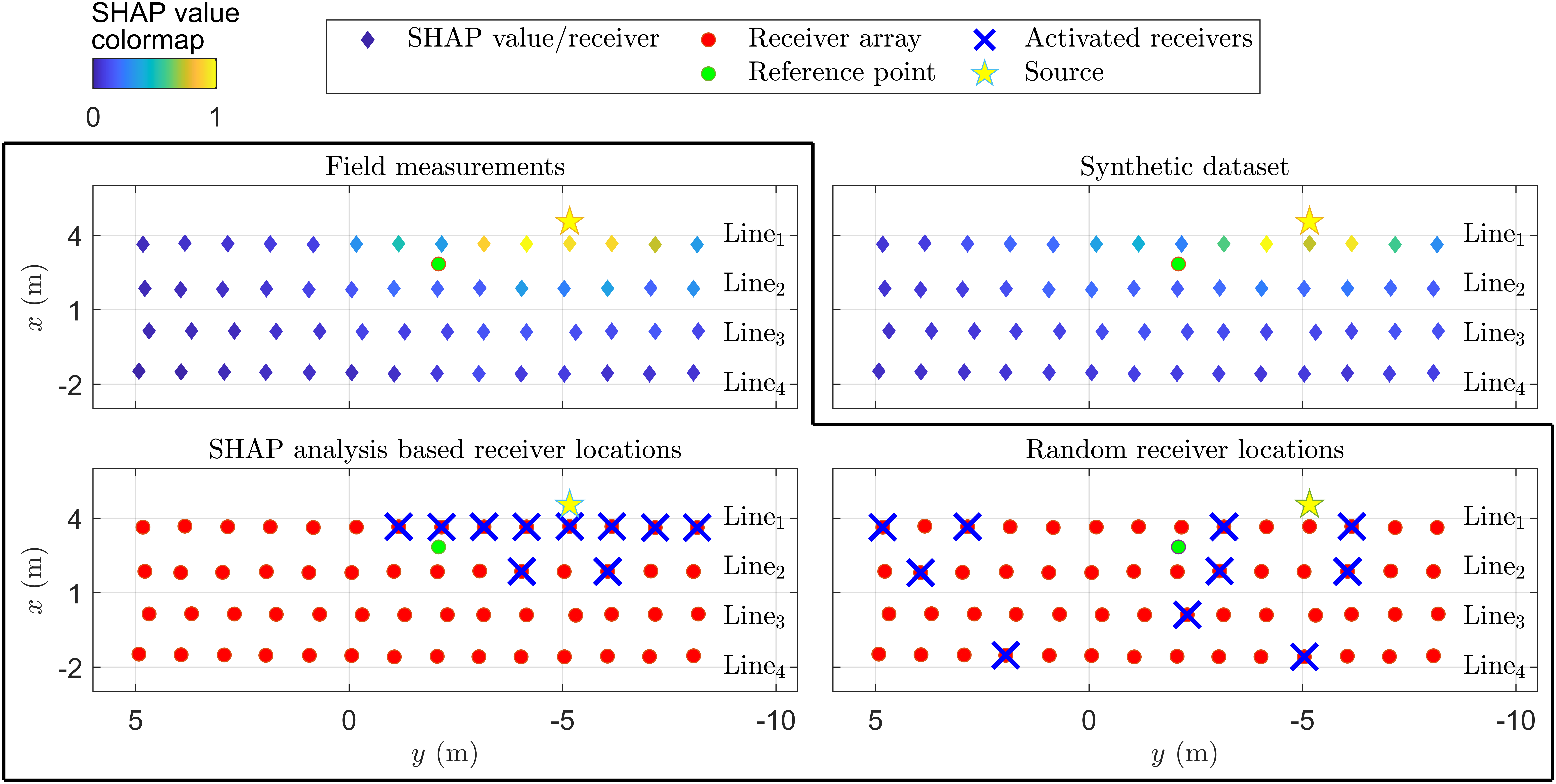}
\centering
\caption{Normalized average absolute Shapley values for predicting water volume from seismic data. The results for the test and real datasets are displayed. Higher Shapley values are correlated with receivers that are closer to the seismic source.}\label{fig:shap}
\end{figure}

10,000 randomly selected samples from the training dataset are used to train the deep explainer model for the SHAP software \cite{Lundberg:2017ua}. The explainer model is then applied to all samples in the field measurements database. After calculating the Shapley values, we compute the normalized mean absolute values for each receiver (see top panel of Fig. \ref{fig:shap}). The results for water volume indicate that the most contributing receivers are those closest to the seismic source. For comparison, we also applied the explainer model to 1,000 randomly selected samples from the synthetic database, revealing a similar distribution of the most contributing receivers.

Next, we constructed two new receiver configurations and trained the neural network model for the field measurements based on SHAP values. For the first configuration, we selected ten receivers having the largest SHAP values, and for the second configuration, we randomly selected ten receivers from the full sensor array (see bottom panel of Fig. \ref{fig:shap}). The KerasTuner optimized network with SHAP analysis-based receiver selection consists three hidden layers with 3,490 (layer 1), 3,780 (layer 2), and 2,580 (layer 3) neurons, a LeakyReLU activation function, a learning rate of 1e-4, and $L_2$ regularization penalty factor of 1e-6. Similarly, the randomly selected receiver selection lead to optimized network with two hidden layers with 3,490 (layer 1) and 2,580 (layer 2) neurons, a LeakyReLU activation function, a learning rate of 1e-4, and $L_2$ regularization penalty factor of 1e-6.

\begin{figure}[!htb]
\includegraphics[height=0.4\textwidth]{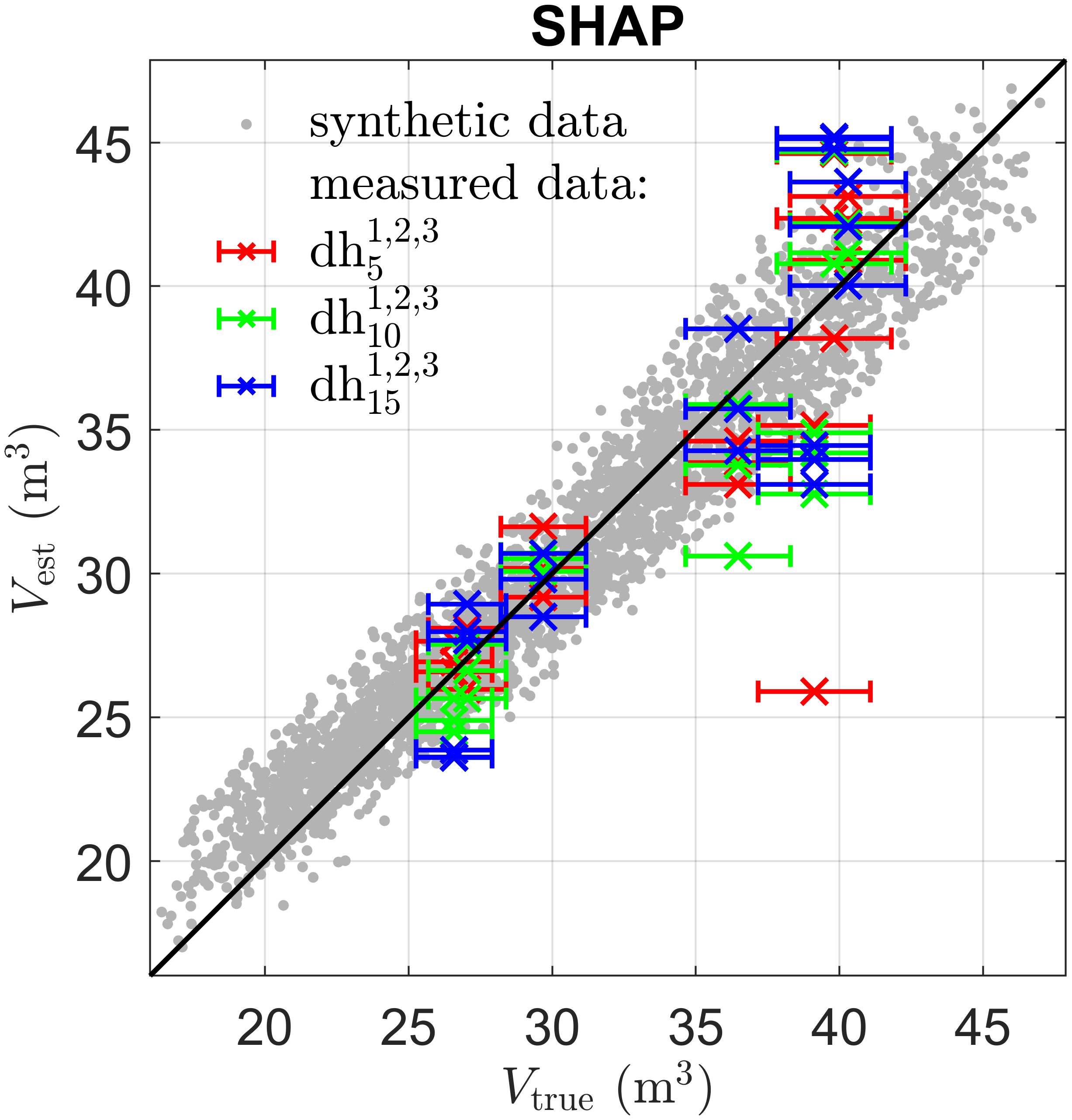}
\includegraphics[height=0.4\textwidth]{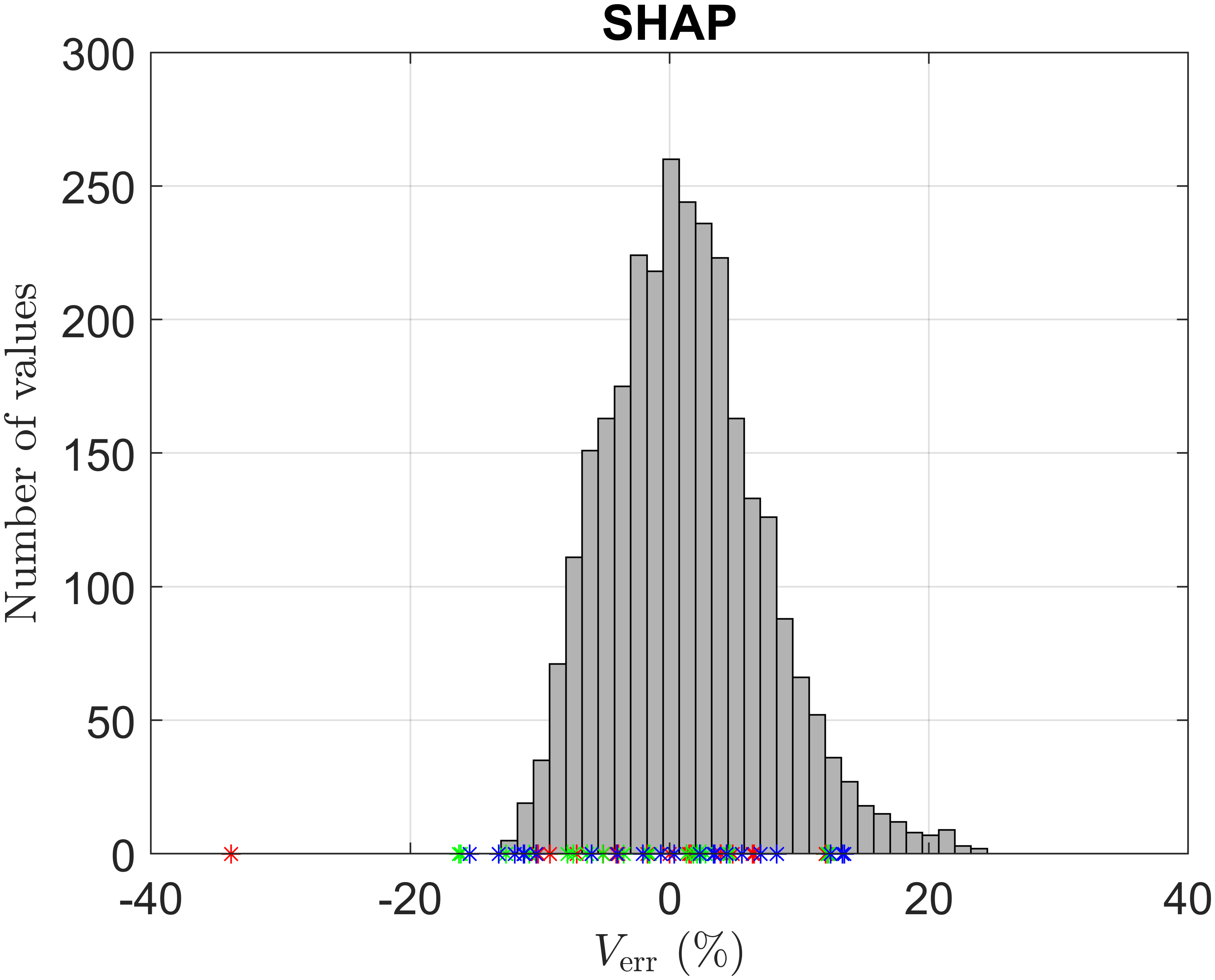}
\includegraphics[height=0.4\textwidth]{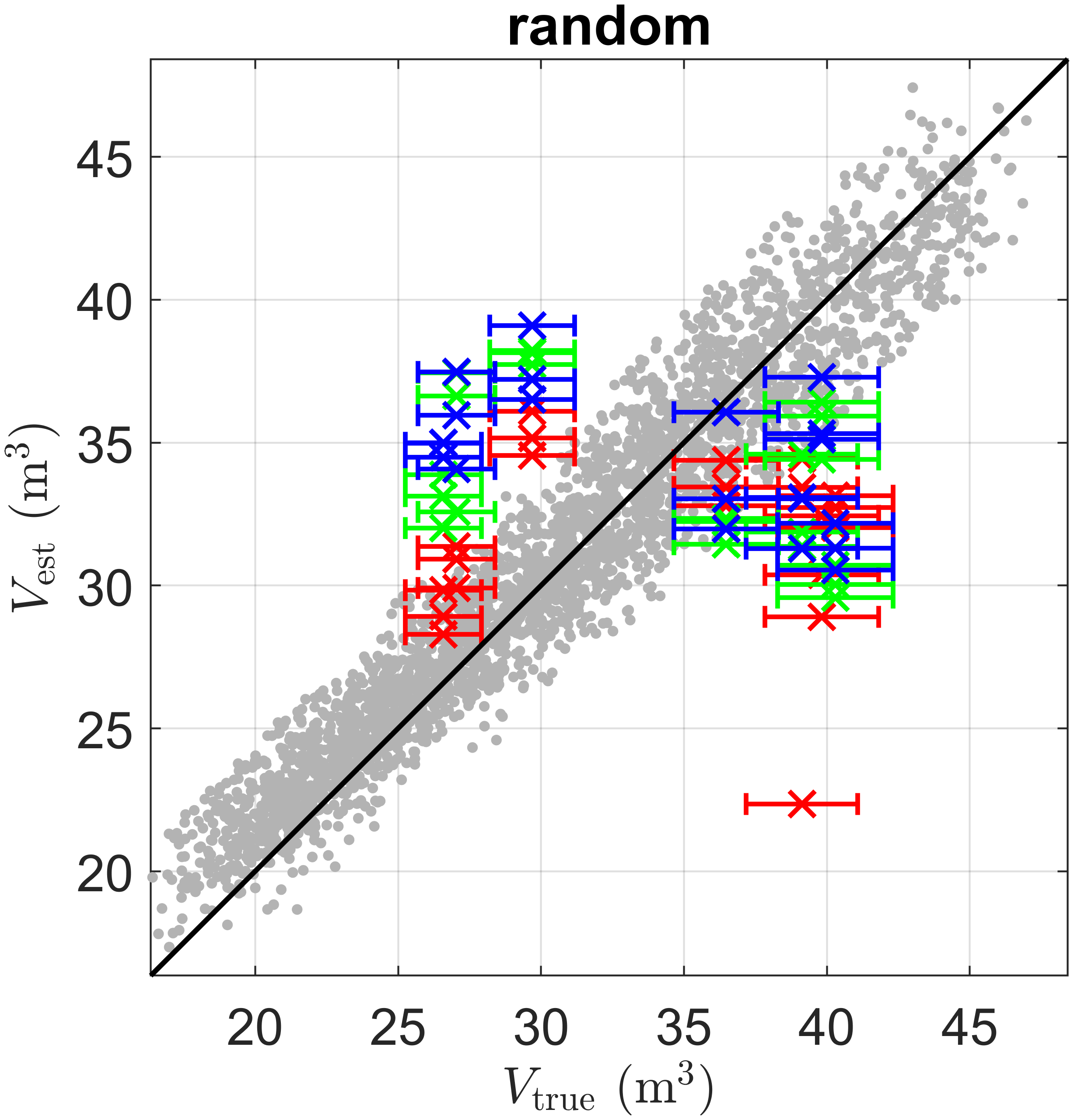}
\includegraphics[height=0.4\textwidth]{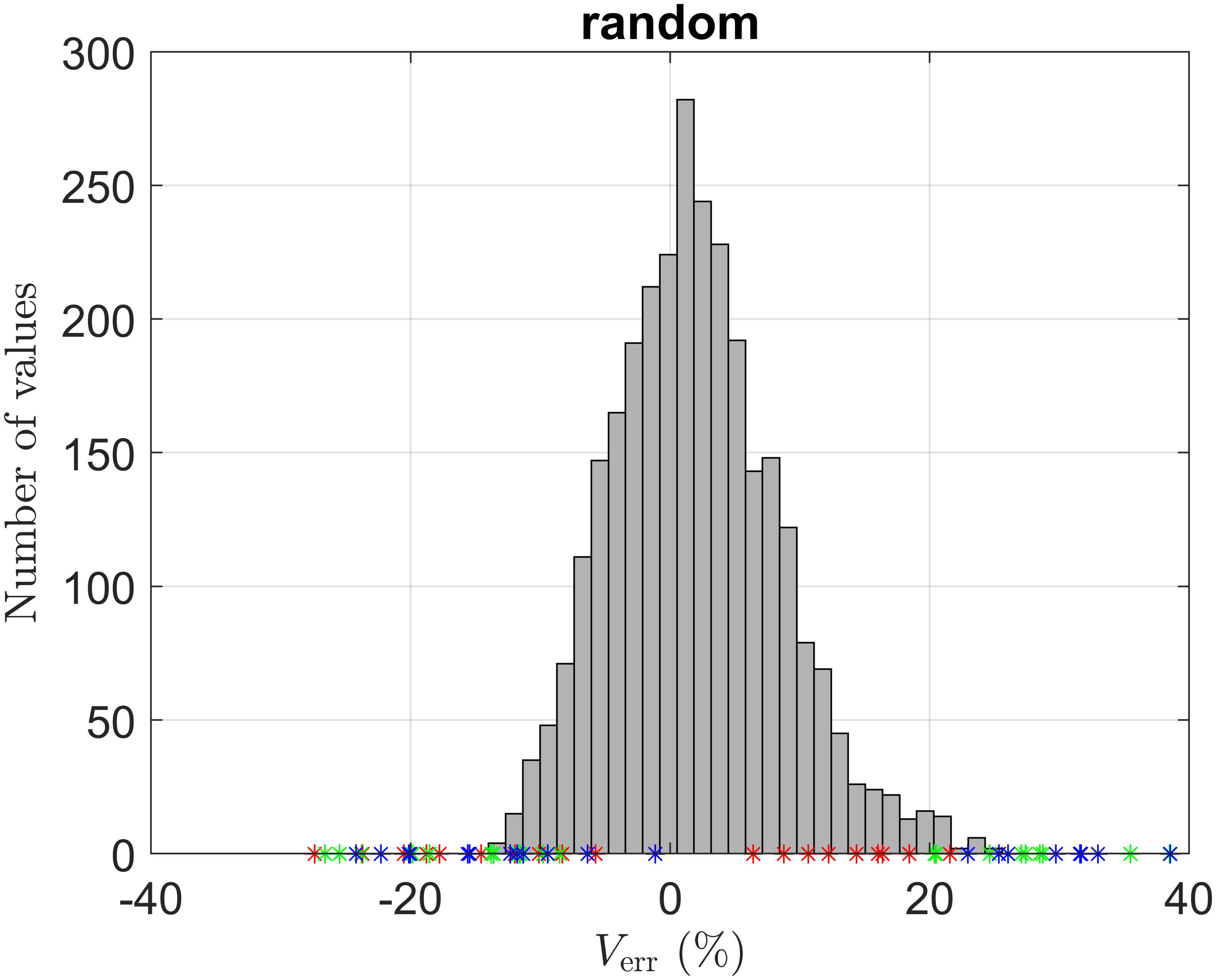}
\centering
\caption{Left: Estimated volumes of water as a function of true value. Right: Histogram of relative prediction errors. Top panel shows the results for receiver configuration based on SHAP analysis and bottom panel for the randomly chosen receivers.  }\label{fig:est_shap}
\end{figure}

Figure \ref{fig:est_shap} displays the estimated water volume as a function of the true water volume for both receiver configurations. The field measurement results reveal a significant impact on estimation accuracy when employing SHAP analysis-based receiver selection compared to random selection. However, with a synthetic database, this effect is not that significant.

The normalized mean biases (NMB), mean absolute errors (MAE), and root mean square errors (RMSE) are used as evaluation metrics to quantitatively analyze the estimation results for all three receiver configurations, see Table \ref{tab:bias}. The table shows that the full receiver array results in the most accurate measures.

\begin{table}[!htb]
    \centering
    \caption{The normalized mean biases (NMB), mean absolute errors (MAE), and root mean square errors (RMSE) were computed for both field measurement and synthetic test databases. It is important to note that, instead of averaging estimates from ten randomly initialized network trainings, they were treated as individual samples in this table. The first two rows show the results for the full receiver array, the following two lines for the ten receivers found most contributing according to SHAP analysis, and the last two rows the results for ten receivers selected randomly.}\label{tab:bias}
    \begin{tabular}{llccc}
\hline
    Receiver array setup & Database & NMB (\%) & MAE (m{}$^3$) & RMSE (m{}$^3$) \\
    \hline
    \multirow{2}{*}{full} & field meas. & 1.2604 & 2.3259 & 2.9787 \\ 
    \cline{2-5}
                         & synthetic & 1.5976 & 1.3801 & 1.7174 \\
    \hline
    \multirow{2}{*}{SHAP} & field meas. & -1.3241 & 3.2449 & 4.1318 \\
    \cline{2-5}
                         & synthetic & 0.5348 & 1.5146 & 1.8950 \\
    \hline
    \multirow{2}{*}{random} & field meas. & -2.3320 & 6.5112 & 7.2178 \\
    \cline{2-5}
                           & synthetic & 1.2344 & 1.5768 & 1.9630 \\
\hline
    \end{tabular}
\end{table}

The evaluation metrics indicate that the full receiver array provides the highest estimation accuracy, while SHAP value-based selection offers a practical compromise with somewhat reduced accuracy. Random receiver selection significantly degrades performance, underscoring the importance of receiver placement and data quality in seismic monitoring. These findings emphasize the value of explainability methods such as SHAP in guiding sensor array design and improving robustness of neural network-based estimations in real-world conditions.

\section{Conclusions}\label{sec:conc}

In this study, we investigated the estimation of water volume in a reservoir using both field and synthetic seismic data. The reservoir’s controllable water-table level and known physical characteristics enabled direct validation of our neural network-based estimation approach. Synthetic training data were generated using a coupled poroviscoelastic–viscoelastic wave propagation model solved with a high-order discontinuous Galerkin method on a GPU cluster. The forward simulations used domain-wise homogeneous material properties and allowed for varying water-table levels within the geometry.

A fully connected neural network was trained to estimate water volume from frequency-domain seismic data normalized by a deconvolution method that accounts for unknown source wavelets. Training data were corrupted with noise levels matching field data conditions, ensuring robustness against realistic measurement noise.

The neural network results indicate that, with a full receiver array, water volume can potentially be estimated with good accuracy, approaching the known reservoir volumes. Analysis using SHAP values suggests that, in the present controlled setup, selecting the top ten most influential receivers can maintain performance comparable to the full array, while random receiver selection tends to degrade accuracy, particularly for field data. These findings highlight that receiver contributions influence estimation accuracy in this experiment. However, they are specific to the present setup and should not be assumed to generalize to other geological settings or problems.

Overall, this work presents a promising framework for seismic-based water volume estimation using neural networks. The findings emphasize the value of carefully chosen receiver configurations and suggest that models trained on synthetic data may generalize to controlled real-world conditions. While this study focuses on a simplified and well-constrained setup, the results point to the potential applicability of this approach in more complex environments in future work.

\section*{Acknowledgements}

This work has been supported by the Research Council of Finland (the
Finnish Centre of Excellence of Inverse Modeling and Imaging),
Flagship of Advanced Mathematics for Sensing Imaging and Modelling
(grant no. 358944), and the Research Council of Finland project
321761. The authors also wish to acknowledge the CSC – IT Center for
Science, Finland, for generously sharing their computational
resources. Special thanks to the Natural Resources Institute Finland
(Luke) for sharing information about the sand pool and allowing us to
carry out measurements on their premises. Authors would also like to
thank Dr. Tuomo Savolainen from the Department of Technical Physics,
University of Eastern Finland, for building the seismic source used in
this work. The 3C nodal receivers used are a part of the Finnish
national pool of seismic instruments \cite{hillers25}.

Last, we would like to acknowledge our colleague, Kai Nyman, who contributed planning and implementation of the Laukaa measurements. Nyman passed away in 2022. 

\bibliographystyle{abbrv}


\appendix
\section{Measured data, water-table level -36.2 cm}

In this section, the data is shown for all measurements from drop height 5 cm at the water-table level -36.2 cm. Notably, the first measurement within this series led to biased water volume estimate, as elaborated in Section \ref{sec:predictions}.

\begin{figure}[!htb]
\includegraphics[width=0.95\textwidth]{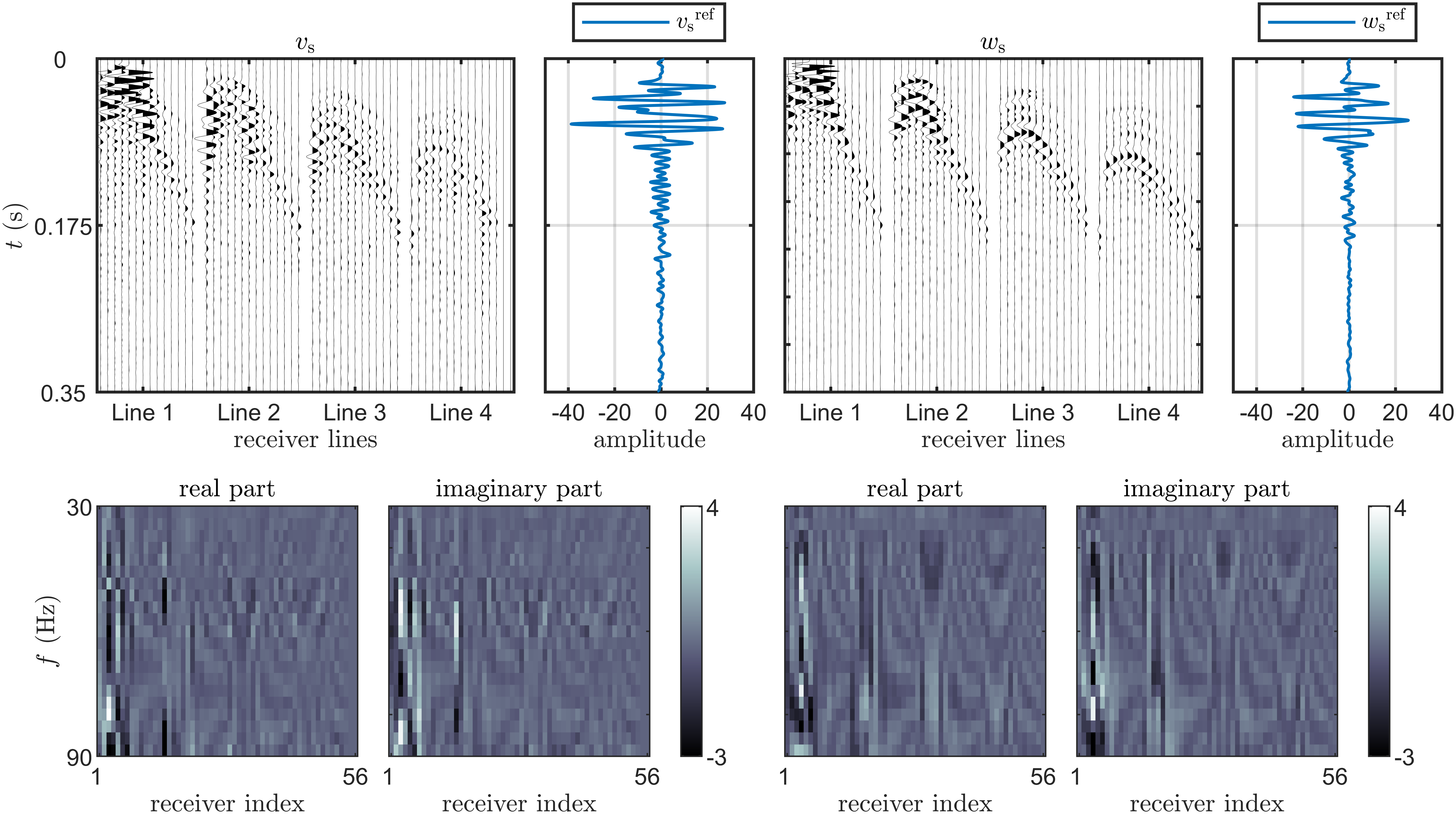}
\centering
\caption{The figure depicts measured data traces, with the top panels displaying results in the time domain and the bottom panels showing the corresponding data converted into the frequency domain. These measurements were taken at a water-table level of -36.2 cm. Specifically, the figure presents the initial measurement out of a series of three, all conducted from a drop height of 5 cm.}\label{fig:est_data71}
\end{figure}

\begin{figure}[!htb]
\includegraphics[width=0.95\textwidth]{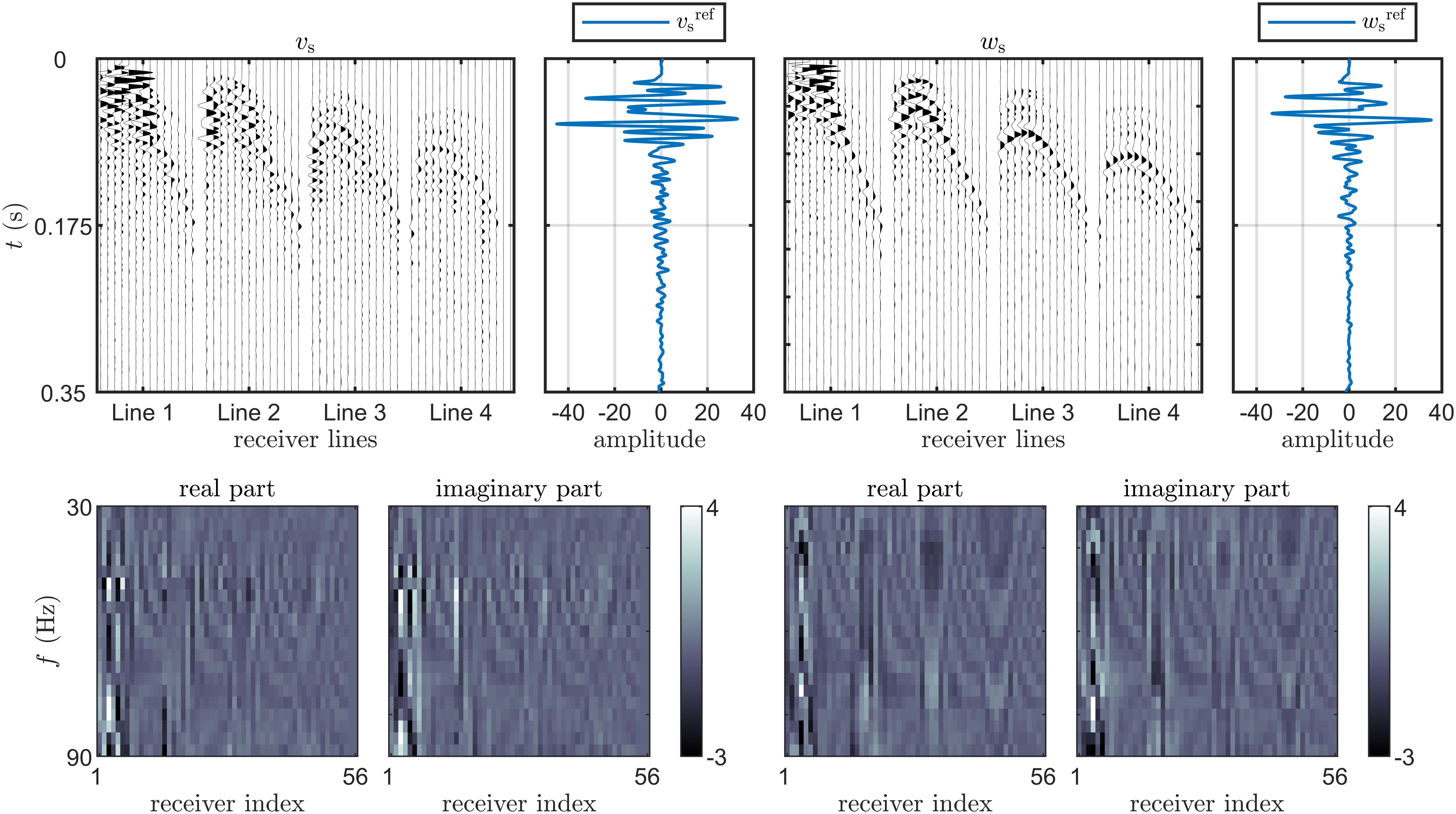}
\centering
\caption{Otherwise the same caption as in Fig. \ref{fig:est_data71}, but the data is for the second iteration of measurements.}\label{fig:est_data72}
\end{figure}

\begin{figure}[!htb]
\includegraphics[width=0.95\textwidth]{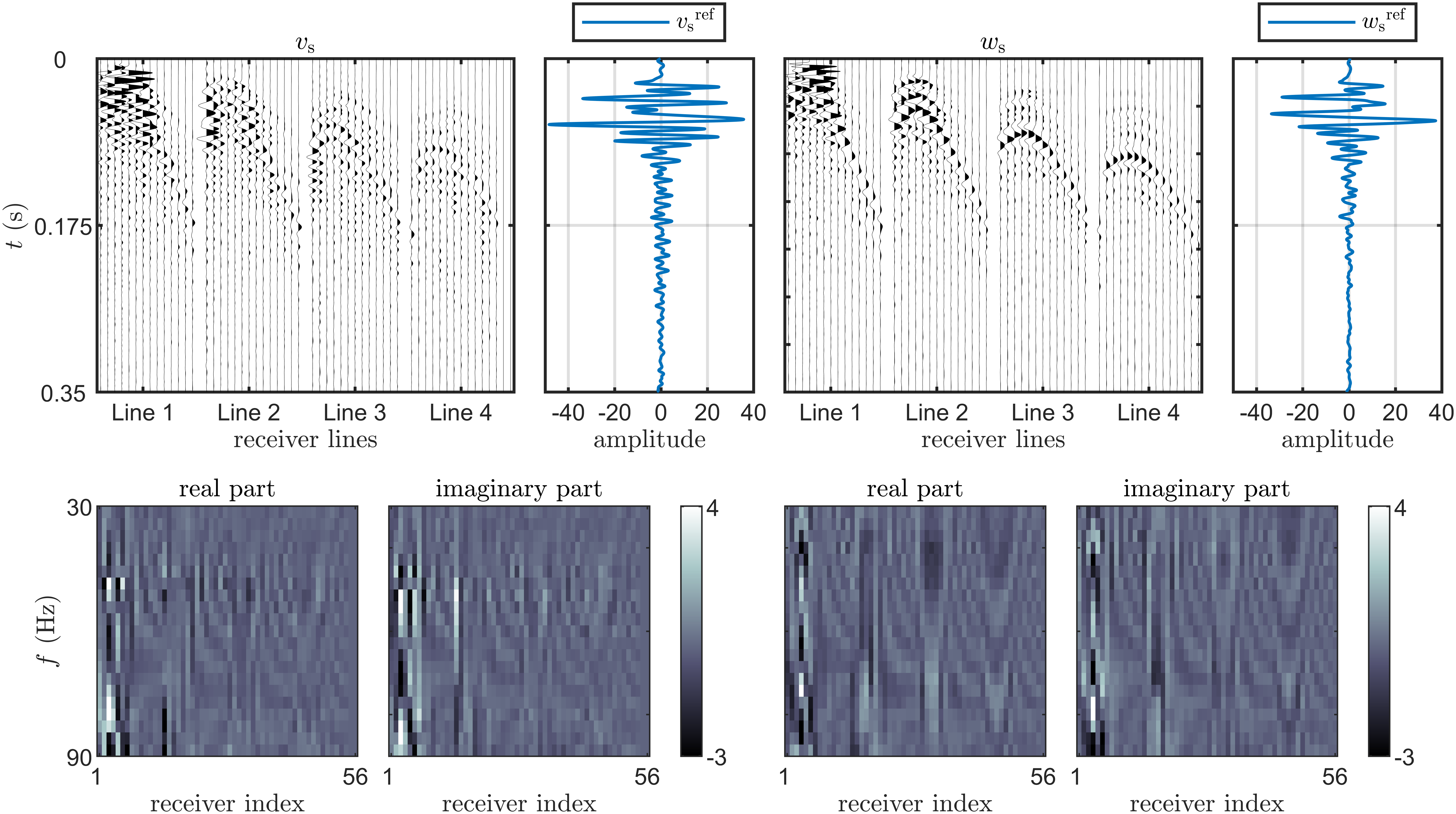}
\centering
\caption{Otherwise the same caption as in Fig. \ref{fig:est_data71}, but the data is for the third iteration of measurements.}\label{fig:est_data73}
\end{figure}

\end{document}